\documentclass[useAMS,usenatbib]{mnras}

\usepackage{times}
\usepackage{graphics,epsfig}
\usepackage{graphicx}
\usepackage{amsmath}
\usepackage{amssymb}
\usepackage{bm}
\usepackage{dcolumn}
\usepackage{epsfig}
\usepackage{subfig}
\usepackage{tabularx}
\usepackage[usenames,dvipsnames,svgnames,table]{xcolor}

\newcommand{\kms}{km~s$^{-1}$}

\newcommand{\ha}{\ensuremath{{\rm H}\alpha}}

\newcommand{\arcs}{\ensuremath{^{\prime\prime}}}

\newcommand{\hicm}{H{\sc i} 21 cm }
\newcommand{\hi}{H{\sc i} }

\title[ESO 184$-$G82: a collisional ring galaxy]{The host galaxy of GRB 980425 / SN1998bw: a collisional ring galaxy
}

\author[Arabsalmani et al.]{M. Arabsalmani$^{1,2}$\thanks{E-mail: marabsalmani@cea.fr}, S. Roychowdhury$^{3,4}$, T. K.  Starkenburg$^{5}$, L. Christensen$^{6}$, 
\newauthor E. Le Floc'h$^{1,2}$, N. Kanekar$^{7}$, F. Bournaud$^{1,2}$,  M. A. Zwaan$^{8}$, J. P. U. Fynbo$^{9}$,  
\newauthor P. M\o ller$^{8}$, E. Pian$^{10}$ \\
       \\
       $^1$ IRFU, CEA, Universit\'e Paris-Saclay, F-91191 Gif-sur-Yvette, France\\
       $^2$ Universit\'e Paris Diderot, AIM, Sorbonne Paris Cit\'e, CEA, CNRS, F-91191 Gif-sur-Yvette, France\\
       $^3$ Jodrell Bank Centre for Astrophysics, School of Physics \& Astronomy, The University of Manchester, Oxford Road, Manchester M13 9PL, UK\\
       $^4$ Institut d'Astrophysique Spatiale, CNRS, Université Paris-Sud, Université Paris-Saclay, Bât. 121, 91405 Orsay Cedex, France \\
       $^5$ Flatiron Institute, 162 Fifth Avenue, New York NY 10010, USA  \\
       $^6$ Dark Cosmology Centre, Niels Bohr Institute, University of Copenhagen, Juliane Maries Vej 30, DK-2100 Copenhagen \O, Denmark\\
       $^7$ National Centre for Radio Astrophysics, Tata Institute of Fundamental Research, Pune University, Pune 411007, India \\
       $^8$ European Southern Observatory, Karl-Schwarzschild-Strasse 2, 85748 Garching bei M\"{u}nchen, Germany\\
       $^9$ Cosmic Dawn Center, Niels Bohr Institute, University of Copenhagen, Juliane Maries Vej 30, DK-2100 Copenhagen $\O$; DTU-Space, Technical \\University of Denmark,  Elektrovej 327, DK-2800 Kgs. Lyngby, Denmark\\
       $^{10}$ INAF, Astrophysics and Space Science Observatory, via P. Gobetti 101, 40129 Bologna, Italy.
}

\begin{document}
\date{}

\pagerange{\pageref{firstpage}--\pageref{lastpage}} \pubyear{}

\maketitle

\label{firstpage}

\begin{abstract}
We report Giant  Metrewave Radio Telescope (GMRT) , Very Large Telescope (VLT) and Spitzer Space Telescope observations of ESO 184$-$G82, the host galaxy of GRB 980425/SN 1998bw, that yield evidence of a companion dwarf galaxy at a projected distance of 13 kpc. 
The companion, hereafter GALJ193510-524947,  is a gas-rich, star-forming galaxy with a star formation rate of  $\rm0.004\,M_{\odot}\, yr^{-1}$, a gas  mass of   $10^{7.1\pm0.1} M_{\odot}$,  and   a stellar mass of  $10^{7.0\pm0.3} M_{\odot}$.  
The interaction between ESO 184$-$G82 and GALJ193510-524947  is   evident from the extended gaseous structure between the two galaxies in the  GMRT \hicm map. 
We  find  a  ring of high column density \hi gas, passing through the actively star forming regions of ESO 184$-$G82 and the GRB location. 
This  ring  lends support to the picture in which ESO 184$-$G82 is interacting with GALJ193510-524947. 
The massive stars in GALJ193510-524947 have similar ages to those in star-forming regions in ESO 184$-$G82, also suggesting that the interaction may have triggered  star formation in both galaxies. 
The gas and star formation properties of ESO 184$-$G82 favour a head-on collision with GALJ193510-524947  rather than a classical tidal interaction.  We  perform state-of-the art simulations of dwarf--dwarf mergers and confirm that the observed properties of ESO 184$-$G82 can be reproduced by  collision with a small companion galaxy.  
This is a very clear  case of  interaction in a gamma ray burst host galaxy, and of interaction-driven star formation giving rise to a gamma ray burst in a dense environment.

\end{abstract}


%-------------------------------------------------------------------------------------------

\section{Introduction}
\label{sec:int}

Long duration Gamma Ray Bursts (GRBs) are luminous   explosions in the Universe, with powerful energy releases that make them detectable back to when the first stars and galaxies were formed \citep[e.g.,][]{Tanvir09}. For a few seconds, these extremely bright explosions release the energy that our Sun emits in its whole lifetime \citep[see][]{Piran13-2013RSPTA.37120273P}.  
Their short durations (a few seconds to minutes) and enormous energy releases can be explained by radiation from   highly relativistic outflowing particles with Lorentz factors $> 100$ \citep[see][and references therein]{Piran04-2004RvMP...76.1143P}. 
Such outflowing jets can be powered by rotational energy tapped from the compact remnants, magnetars or black holes, of the core-collapse of massive stars  \citep[][]{Usov92-1992Natur.357..472U, Woosley93-1993ApJ...405..273W, MacFadyen99-1999ApJ...524..262M,  Aloy00-2000ApJ...531L.119A, Zhang03-2003ApJ...586..356Z, Zhang04-2004ApJ...608..365Z, Yoon05-2005A&A...443..643Y, Woosley06-2006ApJ...637..914W}. 
This is observationally  supported by the occurrence  of GRBs in  actively  star-forming regions, which links GRB formation to  massive stars  \citep[][]{Paczynski98-1998ApJ...494L..45P, Fynbo00, Bloom02-2002AJ....123.1111B, Lefloch03-2003A&A...400..499L, Christensen04-2004A&A...425..913C, Fruchter06-2006Natur.441..463F, Fynbo06-2006Natur.444.1047F,  Lyman17-2017MNRAS.467.1795L}. The association of  GRBs with type Ib,c supernovae (SNe) \citep[e.g.,][]{Galama98-1998Natur.395..670G, Hjorth03-2003Natur.423..847H, Stanek03-2003ApJ...591L..17S, Malesani04-2004ApJ...609L...5M, Pian06-2006Natur.442.1011P}, and also the young ages of stellar populations in GRB  environments \citep[][]{Chary02-2002ApJ...566..229C, Christensen04-2004A&A...425..913C} strengthen the connection between GRBs and massive stars. 

GRBs typically occur in   low mass and metal poor  (dwarf) galaxies    \citep[][]{Fynbo06-2006A&A...451L..47F, Prochaska08-2008ApJ...672...59P, Savaglio09-2009ApJ...691..182S, Castroceron10-2010ApJ...721.1919C, Graham13-2013ApJ...774..119G, Kruhler15-2015A&A...581A.125K, Cucchiara15-2015ApJ...804...51C, Perley16-2016ApJ...817....8P}. 
This is  often interpreted as an indication that a low metallicity is needed  for the formation of GRB progenitors.  Such a hypothesis   is consistent with the    single-star progenitor model for GRB formation where the low metallicity of the progenitor star is critical.  
However,  the detection of several GRBs in metal-rich environments \citep[][]{Savaglio12-2012MNRAS.420..627S, Elliott13-2013A&A...556A..23E, Schady15-2015A&A...579A.126S} and particularly, the identification of a large number of massive and metal rich GRB host galaxies  \citep[dark / dust-obscured GRB hosts,][]{Svensson12-2012MNRAS.421...25S, Hunt14-2014A&A...565A.112H, Perley13-2013ApJ...778..128P} has raised questions about whether a low metallicity is indeed necessary for the formation of GRB progenitor stars.

While the low metallicity requirement   is debated, high star formation densities do appear to play a critical role for the formation of GRB progenitors. 
Studies in both the local and the high-$z$ Universe show that massive stars (and hence GRB progenitors) are more likely to be found in regions  with  high star formation rate (SFR) densities \citep[][]{Dabringhausen09-2009MNRAS.394.1529D, Dabringhausen12-2012ApJ...747...72D, Banerjee12-2012A&A...547A..23B, Marks12-2012MNRAS.422.2246M, Peacock17-2017ApJ...841...28P, Schneider17-2018Sci...359...69S, Zhang18-2018Natur.558..260Z}.  
GRB hosts are indeed found to have   high surface densities of SFR \citep[][]{Kelly14-2014ApJ...789...23K}. 
Interactions are known to  enhance the  star formation activities of galaxies \citep[][]{Renaud14-2014MNRAS.442L..33R} 
and also trigger the formation of massive and compact clumps \citep[see][and references therein]{Renaud-2018NewAR..81....1R}. It would be therefore interesting to investigate whether interactions and mergers are common in GRB host galaxies.

Evidence for interactions is likely to be easier  to obtain in the closest GRB host galaxies,  as their proximity allows us to identify both very faint companion galaxies and  weak disturbances in their velocity fields.  Perhaps the best system in this regard is ESO 184$-$G82, the host of GRB 980425 and its associated supernova, SN 1998bw \citep[][]{Galama98-1998Natur.395..670G}. At $z = 0.0087$, this has the lowest redshift of any known  GRB till date. 
ESO 184$-$G82 is a barred spiral (Sbc-type) galaxy, and has several H{\sc ii} regions that are actively forming stars \citep[][]{Fynbo00}. It has an SFR  of  $0.2-0.4$ $\rm M_{\odot}\, yr^{-1}$ \citep[][]{Christensen08-2008A&A...490...45C, Kruhler17-2017A&A...602A..85K} and a  stellar mass of $\rm 10^{8.7}\,M_{\odot}$ \citep[][]{Michalowski14-2014A&A...562A..70M} which   place  it  on the galaxy Main Sequence relation in the M$_*$--SFR plane \citep[][]{Brinchmann04-2004MNRAS.351.1151B}. 
Most interestingly, there is a very bright star forming region in the host galaxy  with an sSFR  more than an order of magnitude larger than the overall sSFR of the host \citep[][]{Hammer06-2006A&A...454..103H, Christensen08-2008A&A...490...45C}. This is one of the most luminous and infrared-bright H{\sc ii} regions identified to date in the nearby Universe \citep[][]{LeFloch12-2012ApJ...746....7L}.
With a high density of young  and massive Wolf-Rayet (WR) stars with ages less than 3 Myr \citep[][]{Kruhler17-2017A&A...602A..85K}, this WR region  appears to have been formed in a recent episode of star formation \citep[][]{Hammer06-2006A&A...454..103H, LeFloch12-2012ApJ...746....7L}. GRB 980425 occurred in an H{\sc ii} region 800 pc from this WR region. The GRB  H{\sc ii} region contains  young and massive stars  with estimated ages between 5 and 8 Myr \citep[][]{Kruhler17-2017A&A...602A..85K}.

\citet[][]{Fynbo00}  proposed that  interactions could have  triggered   the recent star formation episode in ESO 184$-$G82. However, extensive multi-wavelength studies ESO 184$-$G82 and its surroundings did not yield any sign of interactions or a  companion galaxy. \citet[][]{Christensen08-2008A&A...490...45C} mapped the \ha\, emission from the host and found its  velocity field  to show ordered  rotation without any signature of a disturbance.   
\citet[][]{Foley06-2006A&A...447..891F} studied the field of ESO 184$-$G82 to search for possible companions interacting with the  GRB host, but found all of the observed galaxies in the field to lie at significantly greater distances than ESO 184$-$G82. They concluded that   ESO 184$-$G82 is an isolated dwarf galaxy and interactions  with other galaxies are not responsible for its star formation.

\hicm mapping studies with radio interferometers allow the possibility of tracing the spatial distribution and velocity fields of the neutral hydrogen in nearby galaxies. Such \hicm studies of GRB host galaxies provide a powerful tool to directly test the hypothesis that an interaction might have triggered the star formation that gave rise to the GRB. 
In \citet[][]{Arabsalmani15-2015MNRAS.454L..51A}, we used the Giant Metrewave Radio Telescope (GMRT) to map the \hicm emission from ESO 184$-$G82, finding its gas disc to be disturbed, while the global gas properties of the galaxy appeared similar to those of local dwarfs. This was the first tentative evidence that interactions or a merger event might indeed have played a role in the recent star formation activity of  ESO 184$-$G82.

In this paper we present  deep GMRT H{\sc i} 21 cm emission observations of ESO 184-G82, which allow us to study the structure of the atomic gas in the vicinity of GRB 980425 in detail, with high spatial resolution. We combine our \hicm mapping data with optical and infrared imaging studies to glean further information on the galaxy's star formation history. 
We also use state-of-the-art  simulations to compare the observed properties with a model of a  merger event. 
This paper is organised  as follows. The observations and data reduction are presented in Section \ref{sec:obs}. The \hicm mapping results are discussed in Section \ref{sec:hi}, while Section \ref{sec:sat} provides details on the companion galaxy identified in this paper. Next, Section \ref{sec:interaction} compares the observed \hicm morphology of ESO 182-G82 and its companion to that expected in a  simulation of the merger of two disc galaxies. Sections \ref{sec:dis} and \ref{sec:sum} contain, respectively, a general discussion and a summary of our results.

%-------------------------------------------------------------------------------------------

\section{Observations and data reduction}
\label{sec:obs}

\subsection{H{\sc I} 21 cm emission observations}

We used the L-band receivers of the GMRT to map the \hicm emission of ESO 184$-$G82 on six consecutive days between 2016 March 17 and 2016 March 22 (proposal no: 29\underline{}076; PI: Arabsalmani). The observations used the GMRT Software Backend, with a bandwidth of 4.167 MHz, centred at 1408.246 MHz, and sub-divided into 512 channels, yielding a velocity resolution of 1.7 km~s$^{-1}$ and a total velocity coverage of $\approx 885$~km~s$^{-1}$. 
The southern declination of ESO 184$-$G82 ($\delta \sim -53^\circ$) implies that it is visible from the GMRT for only $\approx 2.5$~hours per day. Our total on-source time from the six runs was hence only 9.6~hours; however, we note that this was significantly larger than the on-source time ($\approx 2.5$~hours) of our earlier run \citep[][]{Arabsalmani15-2015MNRAS.454L..51A}. The bright flux calibrator 3C48 was observed at the start and end of each observing run, to calibrate the system bandpass.

``Classic'' {\sc aips} was used for the analysis of the data \citep[][]{Greisen03-2003ASSL..285..109G}. 
After initial data editing and bandpass calibration, a "channel-0" visibility data set was created by averaging together line-free channels. The flux scale of the data was set by an initial calibration to a sky model based on our earlier GMRT continuum image of the field \citep[][]{Arabsalmani15-2015MNRAS.454L..51A}. This was followed by a standard self-calibration, imaging  and data-editing procedure on the same channel-0 data set, until no further improvement was seen in the continuum image on further self-calibration. The antenna-based gains derived from the above procedure were then applied to all visibilities of the original multi-channel data set. 
At the end of the loop, the final calibration was applied to all the visibilities.

The radio continuum image made using the line-free channels at the end of the self-calibration cycle, was used to subtract the continuum from the calibrated visibilities, using the task {\sc uvsub}.
The residual visibilities were mapped with different U--V tapers to produce spectral cubes at different angular  resolutions using the task {\sc imagr}. 
The velocity resolution was optimized to be 7 \kms\ by averaging groups of four channels together.
This was done to improve the statistical significance of the detected H{\sc i} 21\,cm emission in independent velocity channels while still having sufficient resolution to accurately trace the velocity field. 

The task {\sc momnt} was then applied to the spectral cubes in order to obtain maps of the H{\sc i} total intensity and the intensity-weighted velocity field at different angular resolutions.
{\sc momnt} works by masking out pixels in the spectral data cube which lie below a threshold flux in a secondary data cube created by smoothing the original  cube both spatially and along the velocity axis -- the smoothing ensures that any localized noise peaks are ignored and only emission correlated spatially and in velocity is chosen.
We created the secondary data cube by applying Hanning smoothing across  blocks of three consecutive velocity channels, whereas spatially a Gaussian kernel of full width at half maximum (FWHM) equal to six pixels was applied.
The threshold flux used to select pixels was approximately 1.5 times the noise in a line-free channel of the original cube. 

We produced four spectral cubes with different angular resolutions. 
The synthesized beam FWHMs for these cubes are    3.0\arcs$\times$9.3\arcs, 11.9\arcs$\times$16.8\arcs, 17.2\arcs$\times$23.9\arcs, and 25.3\arcs$\times$44.1\arcs. The properties of the four  cubes are listed in Table~\ref{tab:obs}.

\begin{table}
\caption{Parameters of the GMRT H{\sc i} data cubes used in this paper.}
\label{tab:obs}
\begin{tabular}{cccc}
\hline
Synthesized Beam&Channel width&Noise in line-free channel\\
(arcs$\times$arcs)&(\kms)&(mJy Bm$^{\rm -1}$)\\
\hline
~3.0$\times$~9.3&7.0&0.7\\
11.9$\times$16.8&7.0&1.0\\
17.2$\times$23.9&7.0&1.2\\
25.3$\times$44.1&7.0&1.7\\
\hline
\end{tabular}
\end{table}

\subsection{Ancillary data}

We obtained   ancillary  data for the field of ESO 184$-$G82 at several wavelengths.  
These include (i)  optical broad-band images of the field  obtained with the  FOcal Reducer and low dispersion Spectrograph ({\it FORS}) on the Very Large Telescope ({\it VLT}) on October 10, 14, and 15, 1999,  using the $\rm B_{BESS}$, $\rm V_{BESS}$, $\rm I_{BESS}$, and $\rm R_{BESS}$ filters, with a total exposure time of 300 seconds in each filter  (Program IDs: 064.H-0375(A) and 066.D-0576(A), PI: F. Patat), (ii) a  narrow-band image of the field  with the  $\rm H_{\alpha}$ filter  obtained  with VLT/FORS on August 03, 2000, with a total exposure time of 300 seconds (Program ID: 165.H-0464(A), PI: Van Den Heuvel), (iii) a  4.5 $\mu$m continuum data set obtained with the InfraRed Array Camera (IRAC) on the {\it Spitzer Space Telescope} on 2004 April 03,  with a total  exposure time of 100 seconds  \citep[as a part of the IRS Guaranteed Time,][Program ID:76]{Houck04-2004ApJS..154...18H}, and (iv) a {\it Hubble Space Telescope} ({\it HST}) image  with the MIRVIS/Clear filter centred at 5737.453 \AA\, on June 11, 2000 with a total exposure time of 295 seconds (Program ID: GO-8640, PI: Holland) 

The VLT/FORS data were analysed following the procedure described by \citet[][]{Sollerman05-2005NewA...11..103S}. The details of the IRAC data reduction are presented in \citet[][]{LeFloch06-2006ApJ...642..636L}, while the HST data analysis is described in \citet[][]{Fynbo00}.

%-------------------------------------------------------------------------------------------

\section{The atomic gas in ESO 184$-$G82}
\label{sec:hi}

\begin{figure*}
\begin{center}
\psfig{file=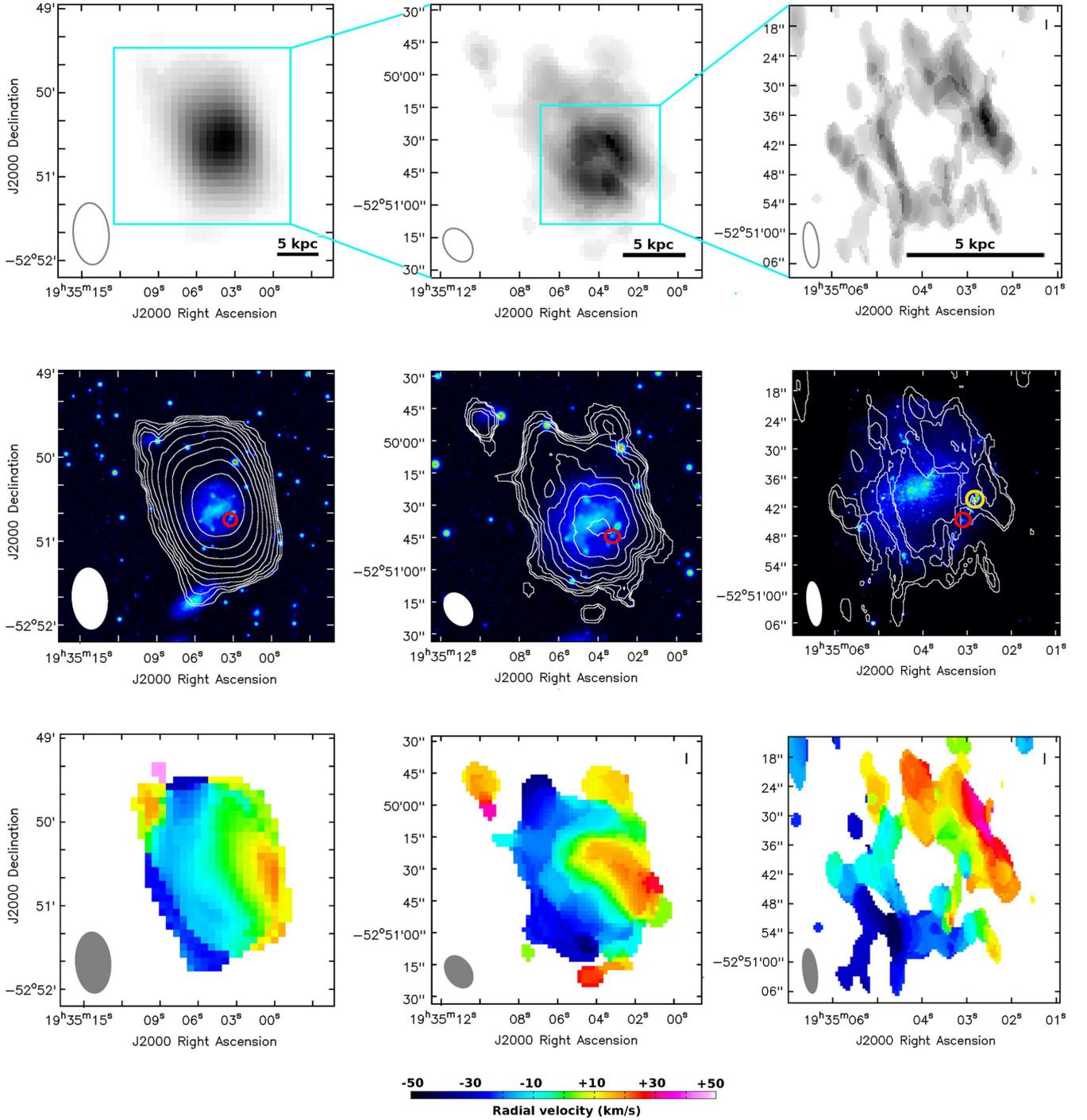,width=7truein}
\end{center}
\caption {{\it {Top row}}: \hicm total intensity maps  in greyscale at three different angular resolutions.  The beam FWHMs (same in all panels of a column) are shown in the bottom-left corner of each panel, and are 25\arcs$\times$44\arcs, 12\arcs$\times$17\arcs, and 3\arcs$\times$9\arcs\  from left to right, respectively. The cyan squares  in the left and middle panels show the area covered in the panel immediately to the right. The black bars represent a physical scale of 5 kpc in each panel. 
{\it {Middle row}}: Contours of the same \hicm total intensity shown in the  top row, overlaid on optical images of ESO 184$-$G82.  The first contour of each \hi intensity map is  at the 3$\sigma$ level of a single channel of the respective data cube. The  first contour is at  $3.6 \times 10^{19} cm^{-2}$(left panel), $1.2 \times 10^{20} cm^{-2}$ (middle panel), and $6.0 \times 10^{20} cm^{-2}$ (right panel), with each subsequent contour in multiples  of $\sqrt{2}$.    
The optical images are the  VLT/FORS B-band image in the left and middle panels,  and  the HST image (MIRVIS filter, centred at 5737.453 \AA) in the right panel. 
In the left panel, the galaxy to the South-East of the GRB host in the FORS image is at $z=0.044$ and hence the extension of gas in that direction is not related to it. 
The  GRB location is marked with the red circles. The  location of the WR region is  marked with a yellow circle in the right panel.   
{\it {Bottom row}}:  \hicm velocity field covering the same spatial area as covered in the respective top and middle  row panels. The colourbar below the middle panel shows the velocity of the gas with respect to the centre of the \hicm emission line, in units of \kms. 
}
\label{fig:ov}
\end{figure*}

\begin{figure*}
\begin{center}
\psfig{file=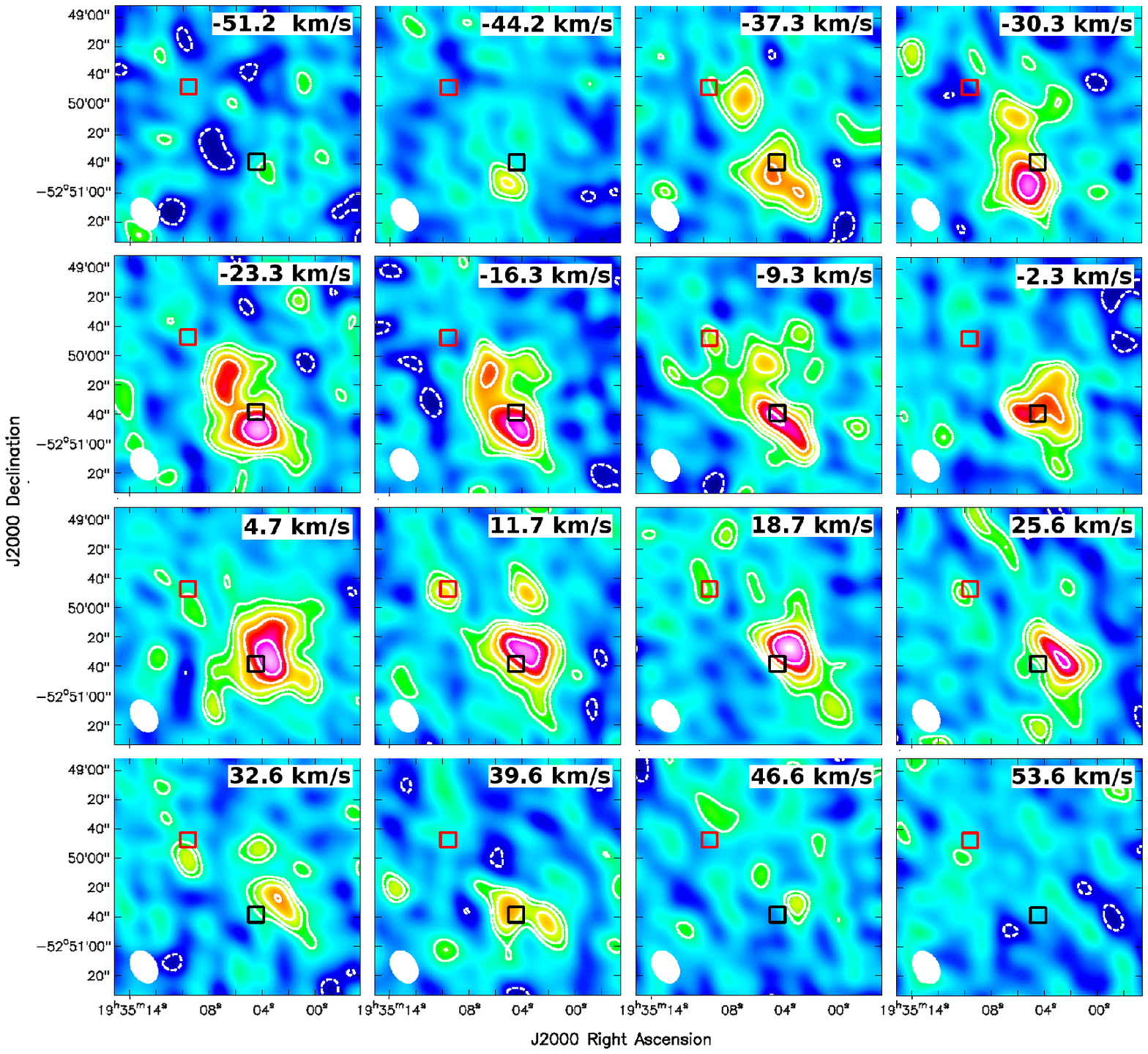,width=1 \textwidth}
\end{center}
\caption{The \hicm flux in successive 7 km~s$^{-1}$ velocity channels of the 17\arcs$\times$24\arcs\  resolution data cube. Contours are overlaid   for clarity. 
The first contour for each channel (positive:  solid, negative: dashed) is at the 2$\sigma$ level, with each subsequent contour in multiples  of $\sqrt{2}$. The beam is shown in the bottom-left corner of each panel. The optical centre of ESO 184$-$G82 and  the centre of the \hi knot are marked with black and red squares, repectivly.}
\label{fig:chans}
\end{figure*}

In Fig. \ref{fig:ov}, we show the \hicm total intensity and velocity maps of ESO 184$-$G82 at three different angular resolutions. 
The synthesized beams are shown in the bottom-left corner of each panel,  and have the FWHMs of   25\arcs$\times$44\arcs, 12\arcs$\times$17\arcs, and 3\arcs$\times$9\arcs for the left, middle, and right  panels respectively (see Table \ref{tab:obs} for the properties  of the \hi cubes). 

The \hicm intensity map at our lowest resolution  (25\arcs$\times$44\arcs) is shown in greyscale in the top-left panel. The corresponding contours are  overlaid on the optical VLT/FORS (B-band) image of the galaxy in the middle-left panel. These maps show the extent of the diffuse gas since we are sensitive to low \hi column densities, $\approx 3.6 \times 10^{19} cm^{-2}$, at this resolution. 
We clearly see  that the diffuse \hi is much more extended than the optical disc of the galaxy (at least twice as large in diameter). 
We derive a total H{\sc i} mass of $\rm 10^{8.90 \pm 0.04}\, M_{\odot}$ for the main gas disc  from the total H{\sc i} flux measured at this  resolution, consistent with the value we reported in \citet[][]{Arabsalmani15-2015MNRAS.454L..51A}. This is comparable to the stellar mass of the galaxy, $M_* = \rm 10^{8.7}\, M_{\odot}$. 
The velocity field at the same coarse resolution (bottom-left panel) shows that the atomic gas disc has ordered rotation, but that the gas to the North-East corner seems not to be following the rotation of the gas disc.

The \hi intensity map  at a resolution of 12\arcs$\times$17\arcs\, in greyscale, and with  contours   overlaid on the  VLT/FORS B-band image, are shown in the middle panels of the top two rows.  
From these  maps,  we find that, in addition to the main mass of gas coincident with the optical disc of ESO 184$-$G82, the gas disc extends at least 5 kpc to the North of the optical disc.  
We clearly detect the presence of an H{\sc i} knot to the North-East of the galaxy, about an arcminute  from the optical centre of ESO 184$-$G82 (middle panels). The extension of gas  
towards this knot is  strong evidence of tidal interaction between ESO 184$-$G82 and the object associated with the \hi knot.
The velocity field at the same  resolution (bottom-middle panel) shows that whereas the main gas disc appears to have regular rotation along an axis running through the centre of the optical disc from South-East to North-West, the extension to the North contains kinematically disturbed gas. The presence of disturbed gas  strengthens the case for an ongoing interaction between the GRB host and a companion galaxy. 

The highest resolution \hi map, with a resolution of 3\arcs$\times$9\arcs, is shown in greyscale in the top-right panel. The corresponding contours, overlaid on the HST image of the galaxy, are shown in the middle-right panel. At this resolution we are sensitive to only high \hi column density  gas, with N(H{\sc i}) $\rm\gtrsim10^{20.8}\,cm^{-2}$.  We find  the  high column density H{\sc i} to have formed  a ring around the optical centre of the galaxy. This ring passes across  the actively star-forming regions in the galaxy and encircles the stellar bar in ESO 184$-$G82. The locations of both the SN/GRB  and the WR region are situated in the Western part of this high column density ring, portions of which were also picked up in our previous GMRT \hicm image \citep[][]{Arabsalmani15-2015MNRAS.454L..51A}. Gas rings can form  due to resonances  with  bars or resonances caused by a mild tidal interaction with a companion galaxy    \citep[resonance rings,][]{Buta96-1996FCPh...17...95B, Buta99-1999Ap&SS.269...79B}. 
But  high density gas rings, which are also the sites of enhanced star formation  in galaxies,   are usually  formed due to collisions  with small companions  \citep[see][for a review on collisional ring galaxies]{Appleton96-1996FCPh...16..111A}.  In  Section \ref{sec:interaction} we discuss  the likely cause for the formation of the high column density gas ring in ESO 184$-$G82.   
The high resolution velocity field (bottom-right panel) shows  that this high column density ring of gas follows the rotation of the main \hi disc of the galaxy, though within the ring there are regions  with velocity gradients  as high as 40 \kms over sub-kpc scales.

In order to take a detailed look at the velocity distribution of the \hi gas, we use the spectral cube with an angular resolution of 17\arcs$\times$24\arcs. 
This intermediate resolution allows us to both be sensitive to relatively low gas column densities and spatially distinguish the \hi knot to the North-East from the main gas disc of ESO 184$-$G82.
Fig. \ref{fig:chans} shows the H{\sc i} fluxes per 7 \kms\  velocity channel at this resolution.
There is H{\sc i} around the optical centre of the galaxy which shows ordered rotation -- the peak emission shifts from South-East to North-West with increasing velocity.
But there appears to be a substantial amount of H{\sc i} which does not follow the ordered rotation, located to the North--North-East of the optical centre.
Emission from the previously mentioned spatially distinct North-Eastern knot is detected in the velocity channels between 2591.7 \kms\ and 2612.6 \kms.
Note that much of the kinematically disturbed gas in various velocity channels is extended towards the location of this North-Eastern knot, reminiscent of gas being dragged out of a galactic halo by the  passage of another galaxy during a merger event.

%-------------------------------------------------------------------------------------------
\begin{figure*}
\begin{center}
\psfig{file=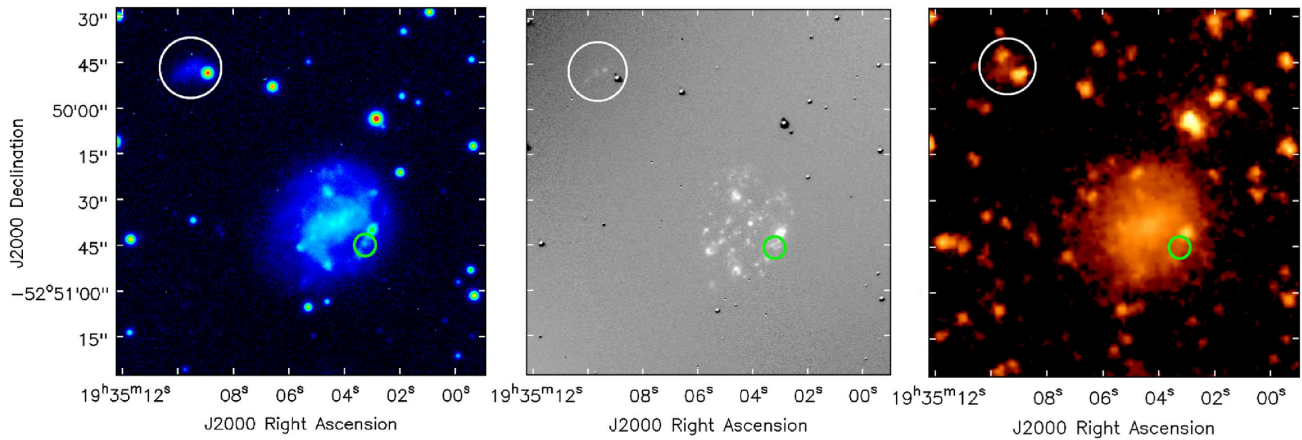,width=0.98\textwidth}
\end{center}
\caption{Ancillary observations of the field around ESO 184$-$G82 show a companion  galaxy co-spatial with the North-Eastern H{\sc i} knot (at a projected distance of 13 kpc from the centre of ESO 184$-$G82),  marked with a white circle in each of the panels. We name this newly identified galaxy  GALJ193510-524947.  Panels from left to right: FORS B band, VLT/FORS narrow-band image centred on \ha, Spitzer/IRAC 4.5 $\mu$m. Note that there is a bright star in the foreground of GALJ193510-524947 in the three images, with a second NIR-bright foreground star in the Spitzer image. The  GRB location is marked with the green circles.}
\label{fig:anc}
\end{figure*}

\section{The companion  galaxy of ESO 184$-$G82}
\label{sec:sat}

The structure of the atomic gas in ESO 184$-$G82 shows clear evidence for an ongoing interaction between the galaxy and a companion object, probably associated with the \hi knot of Fig.~\ref{fig:ov}. 
In order to identify the  optical counterpart of the  H{\sc i} knot, we searched the multiple VLT/FORS optical images. In all these images we  clearly detect   optical emission spatially coincident with the H{\sc i} knot. We also identify this object  in  the Spitzer/IRAC2 4.5~$\mu$m image and the VLT/FORS H$\alpha$ image.  The optical and infrared images of the field are shown in Fig.~\ref{fig:anc} (see also the top-middle panel of Fig. \ref{fig:ov}).   
These confirm  the association of the \hi knot with a  galaxy centred on $\rm RA = 19h 35m 09.6s$ and $\rm Dec = -52h 49m 46.9s$ and at a projected distance of 13 kpc from the centre of ESO 184$-$G82.   
The \hicm emission from this  galaxy  is centred at a redshift of $z = 0.00867\pm0.00002$, consistent with the  redshift of ESO 184$-$G82 ($z=0.00860\pm0.00002$ obtained from the \hicm emission line); ie.,  the centres of the \hicm emission from the two galaxies  are separated by about 20 \kms. We name this newly identified galaxy  GALJ193510-524947.

GALJ193510-524947 appears to be a star-forming dwarf galaxy. We obtain an H$\alpha$ emission flux of $\rm 3.4 \times 10^{-15} erg\,s^{-1}\,cm^{-2}$, implying that the galaxy has an SFR of $0.004$ $\rm M_{\odot}\, yr^{-1}$  \citep[based on the calibration of][]{Kennicutt98-1998ApJ...498..541K}. 
The VLT/FORS images yield AB magnitudes of B=$18.84 \pm 0.10$, V=$18.33 \pm 0.10$, R=$18.05 \pm 0.15$, and I=$17.93 \pm 0.20$ for the galaxy (corresponding to absolute magnitudes of $-13.66$, $-14.17$, $-14.45$, and $-14.75$, respectively). The Spitzer/IRAC2 photometry is complicated by contamination from two foreground stars in the 4.5$\mu$m image (see the right panel of Fig. \ref{fig:anc}). We hence first measured the contributions of the two stars using aperture photometry, and subtracted their emission to obtain the  flux density of GALJ193510-524.947. This yielded a $4.5\mu$m flux density of $17 \pm 7 \mu$Jy, equivalent to an AB magnitude of $20.8^{+0.6}_{-0.4}$. 
Note that while GALJ193510-524947 is clearly  detected in the Spitzer/IRAC2 image, the presence of the two bright stars in the image lead to  the large uncertainty in the IRAC photometry.

We estimated the stellar mass of GALJ193510-524947 by  modelling its spectral energy distribution (SED) with \textit{LePhare} \citep[][]{1999MNRAS.310..540A} based on our optical and near-infrared (NIR) photometry.  We  use  the  Stellar Population  Synthesis  templates developed by \citet{Bruzual03-2003MNRAS.344.1000B}, assume a Chabrier initial mass function  \citep[][]{2003PASP..115..763C}, and consider  an exponentially declining star formation history (SFR  $\propto {\rm e}^{-t/\tau}$). 
From this, we estimate a  stellar mass of $\rm 10^{7.0 \pm 0.3}\,M_{\odot}$ for GALJ193510-524947. 
We note that our SED modelling is dominated by the optical  photometry which has lower errors than the $4.5\mu$m photometry. 
In order to confirm the stellar mass derived from the SED modelling, we obtain an independent  estimate of the stellar mass of GALJ193510-524947 from its NIR photometry alone.   
For this we assume a ratio of 0.6 for the Spitzer/IRAC 3.6 $\mu$m to 4.5 $\mu$m fluxes  \citep{Zhu10-2010RAA....10..329Z}  and use  the calibration of \citet{Leroy08-2008AJ....136.2782L}.  We then obtain a stellar mass of  $\sim 10^{6.7} M_{\odot}$  from the NIR photometry  which is consistent with our estimate from the SED modelling. 

We  measure an   H{\sc i} mass  of $10^{7.1\pm0.1} M_{\odot}$ for GALJ193510-524947. This is  comparable to its stellar mass,  implying that GALJ193510-524947 is a   gas-rich galaxy.  We detect the \hicm emission line from GALJ193510-524947 in five channels, corresponding to a velocity width of  $\approx 35$ \kms\, (see Fig. \ref{fig:chans}). This places the galaxy on the  baryonic Tully--Fisher relation for low-mass dwarfs in the local Universe \citep[][]{McGaugh12-2012AJ....143...40M}. 
Based on our SED modelling, we derive the continuum emission of GALJ193510-524947  at the rest-frame \ha\, wavelength (6562.8 \AA)  to be $\rm 1.2 \times 10^{-16} erg\,s^{-1}\,cm^{-2}$\AA$^{-1}$. With this and  the \ha\, flux measured from the VLT/FORS narrow-band image we estimate the equivalent width of the \ha\, line (EW$_{H\alpha}$, the ratio of \ha\, flux to the continuum level at the wavelength of the \ha\, emission) to be $\sim$ 28 \AA\, for GALJ193510-524947. Using  Starburst99 models  with instantaneous star formation laws \citep{Leitherer99-1999ApJS..123....3L},  this corresponds to an age of $\sim$ 8 Myr, somewhat larger than, but comparable to, those of the H{\sc ii} regions in ESO 184$-$G82 \citep[see][]{Christensen08-2008A&A...490...45C, Kruhler17-2017A&A...602A..85K}.

\begin{table*}
\caption{Properties of GALJ193510-524947, the companion  of ESO 184$-$G82.}
\label{tab:sat}
\begin{tabular}{ccccccccc}
\hline
B & V & R & I & F$_{{\rm 4.5}\mu{\rm m}}$ & $\rm F_{H\alpha}$     & SFR & $M_*$ & $M_{HI}$\\
  &   &   &   &     $(\mu$Jy)       &($erg\,s^{-1}\,cm^{-2}$) & ($\rm M_{\odot}\, yr^{-1}$) & ($\rm M_{\odot}$) & ($\rm M_{\odot}$)\\
\hline
&&&&&&&&\\
18.84$\pm$0.10 & 18.33$\pm$0.10 & 18.05$\pm$0.15 & 17.93$\pm$0.20 & 17$\pm$7 & $3.4 \times 10^{-15}$ & 0.004 & $10^{7.0\pm0.3}$ & $10^{7.1\pm0.1}$\\
&&&&&&&&\\
\hline
\end{tabular}
\end{table*}

%-------------------------------------------------------------------------------------------

\section{Interaction between ESO 184$-$G82 and GALJ193510-524947}
\label{sec:interaction}

The structure of atomic gas and also the star formation activity in ESO 184$-$G82 and GALJ193510-524947  are reminiscent of an interacting system. In addition to the main   mass of gas which is spatially    coincident with the optical disc of ESO 184$-$G82,  the atomic gas in ESO 184$-$G82  has  an extension over at least 5 kpc (projected size) to the North of the optical disc.  
The eastern part of the extended gas  looks like a bridge between   ESO 184$-$G82 and GALJ193510-524947,  suggestive of a tidal interaction or a collision between  the two galaxies.  
While the  main mass of gas in ESO 184$-$G82 shows regular rotation, a substantial amount of \hi gas  in the extension  is disturbed and does not follow the ordered rotation.  
This is reminiscent of gas being dragged out of ESO 184$-$G82   by the  passage of GALJ193510-524947. 
The \hi gas with the highest  column density  appears to have formed  a ring around the optical centre of ESO 184$-$G82, passing across its actively star-forming regions. This structure, resembling a cartwheel-like ring,  is suggestive of a collisional ring formed due to a head-on collision with a small galaxy like GALJ193510-524947 which passed  through the disk  of ESO 184$-$G82 close to its centre  \citep[a drop-through collision, see][]{Wong06-2006MNRAS.370.1607W}. The large velocity gradients  of the atomic gas  in the  gas ring of ESO 184$-$G82 and also the  presence of the large \ha\, knots coincident with the \hi ring  lend support to this hypothesis.

In a drop-through collision,  the inner annular shells of  gas will have larger  velocities compared to the  outer shells. The catching up of the inner shells  with the outer shells creates   shock and compression of gas and leads to the formation of a ring of high density gas with a large velocity dispersion \citep[][]{Appleton96-1996FCPh...16..111A, Bournaud03-2003A&A...401..817B}. The high velocity dispersion of  gas in the ring  increases the Jeans mass which leads to the  formation of massive molecular gas clumps and hence large knots of star formation \citep[][]{Horellou01-2001Ap&SS.276.1141H, Renaud18-2018MNRAS.473..585R}.   Unlike in classical tidal interactions,  the lack of large gas inflows towards the galaxy centre   during a collision    results in the absence (or negligible amount) of star formation enhancement  in the central regions \citep[][]{Renaud18-2018MNRAS.473..585R}. 
This picture is consistent with the star  formation activity in ESO 184$-$G82.  As is clear from  its  \ha\, emission line observations, the recent star formation  in ESO 184$-$G82 is not concentrated in the centre of the galaxy, but arises in a number of H{\sc ii} regions coincident with the high column density  \hi ring  \citep[see the middle panel of Fig. \ref{fig:anc}; see also][]{Christensen04-2004A&A...425..913C, Kruhler17-2017A&A...602A..85K}.  Moreover,  the velocity map presented in the bottom-right panel of Fig. \ref{fig:ov} clearly shows  large velocity gradients (as large as 40 \kms on sub-kpc scales) along the radius of the high column density gas ring, typical of collisional rings. The gas ring in  ESO 184$-$G82 is also asymmetric, with its centre  offset from the optical centre of the galaxy. It also has  higher column densities in its North-West region compared to those in the rest of the ring. These asymmetric features  too are typical of collisional rings.

The identification of GALJ193510-524947, a companion galaxy associated with the \hi knot to the North-East of ESO 184$-$G82, and at a projected distance of 13 kpc from the GRB host, strengthens  the case for an interaction between the two galaxies.  
We use state-of-the art simulations of dwarf--dwarf mergers \citep[][]{Starkenburg16-2016A&A...587A..24S}  to test whether  the observed gas and star formation properties of  ESO 184$-$G82 and its companion can be reproduced by a dwarf-dwarf interaction model. 
These  are  controlled (isolated) simulations, performed with the OWLS version \citep[][]{Schaye10-2010MNRAS.402.1536S} of the N-body/Smoothed-Particle-Hydrodynamic-code Gadget-3 \citep[][]{Springel01-2001NewA....6...79S, Springel05-2005MNRAS.364.1105S}. 
In these simulations, both dwarf galaxies have a Hernquist dark matter halo, and exponential stellar and gas discs. 
The primary dwarf contains  $5\times10^6$ particles in its dark matter halo and $10^6$ particles  in its baryonic matter. 
The secondary dwarf contains  $10^6$ particles in its dark matter halo, and $2\times10^5$ particles in its  baryons. 
The gravitational softening length is 10 pc for dark matter, and 3 pc for baryonic particles, and the  smoothing is done over 48 neighbors.  
Gas above a density of $0.1$ cm$^{-3}$ is governed by an effective equation of state and  forms stars following a Kennicutt-Schmidt relation while at lower densities it follows an isothermal equation of state \citep[see][]{Schaye08-2008MNRAS.383.1210S}. 
We consider both the non-star-forming gas with temperature $< 2\times10^4$ K, and  the least dense three-fourths of the star-forming gas to be in the atomic phase  \citep[following][]{Genel14-2014MNRAS.445..175G} and assume a mass ratio of molecular gas to atomic gas of $\approx 1/3$ \citep[][]{Saintonge11-2011MNRAS.415...32S}.
Feedback is implemented based on the kinetic stellar wind prescription of \citet{DallaVecchia08-2008MNRAS.387.1431D} and is calibrated to ensure self-regulated star formation.  
For a detailed description of the simulations we refer the readers to \citet[][]{Starkenburg16-2016A&A...587A..24S}.

Our main criteria  are to simultaneously reproduce    the following observed features  of  ESO 184$-$G82: 
(i) the extension of \hi gas from ESO 184$-$G82 towards GALJ193510-524947, (ii) the asymmetric ring of high column density \hi gas in ESO 184$-$G82, (iii) the large velocity gradients of the \hi in the ring, (iv) the enhancement of  star formation in the location of the gas ring, and finally (v) the lack of  enhanced star formation activity   in the central regions  of ESO 184$-$G82.   
We run two sets  of simulations, one for classical tidal  interactions and the other for collisional interactions.   In both we start   with two discs with smoothly  distributed stellar and gas components. In the  tidal interaction simulations,  the secondary  galaxy passes through the outskirts of the primary galaxy disc at the first pericentre,  on an orbit which has a small inclination with respect to the plane of the disc of the primary dwarf (covering a range of -10 to 30 degrees). In the collisional interaction simulations, the secondary  galaxy has a direct collision with the primary galaxy  with  angles between 70 to 90 degrees with respect to  the plane of the primary galaxy.

We find that the  tidal interaction simulations, within the ranges of  the initial configurations mentioned below,  fail to reproduce  the observed criteria (ii) to (v) mentioned above. The explored initial configurations for this set of simulations are:  a virial mass range of  $\rm 5.0 \times 10^{10} - 1.0 \times 10^{11} M_{\odot}$ (in 3 steps) for the  primary dwarf and   $\rm 8.0 \times 10^{8} - 5.0 \times 10^{10} M_{\odot}$ (in 7 steps)  for   the secondary  dwarf galaxy;  
the range of 0.001$-$0.02 (in 5 steps) for the ratio of baryonic to virial mass, 0.3$-$0.7 (in 5 steps) for the ratio of gas to baryonic mass, and 1$-$4 (in 5 steps) for the ratio of gas to stellar disc size for both the dwarf galaxies; the range of -10 to 30 degrees (in 6 steps) for the inclination of the orbit of the secondary galaxy with respect to the disc of the primary dwarf; and the range of 28 to 66 kpc (in 3 steps) for the initial separation between the two dwarf galaxies. 
In particular, in  tidal interactions the star formation enhancement in the central regions of the galaxy (central kpc) should   contribute significantly (if not dominantly) to the total star formation enhancement of the galaxy \citep[][]{Barnes91-1991ApJ...370L..65B, DiMatteo07-2007A&A...468...61D, DiMatteo08-2008A&A...492...31D, Teyssier10-2010ApJ...720L.149T, Powell13-2013MNRAS.434.1028P, Renaud14-2014MNRAS.442L..33R, Hibbard96-1996AJ....111..655H}. 
The fact that this is not the case in ESO 184$-$G82 suggests that it is unlikely that a tidal interaction has taken place in the system.

\begin{figure*}
\begin{center}
\psfig{file=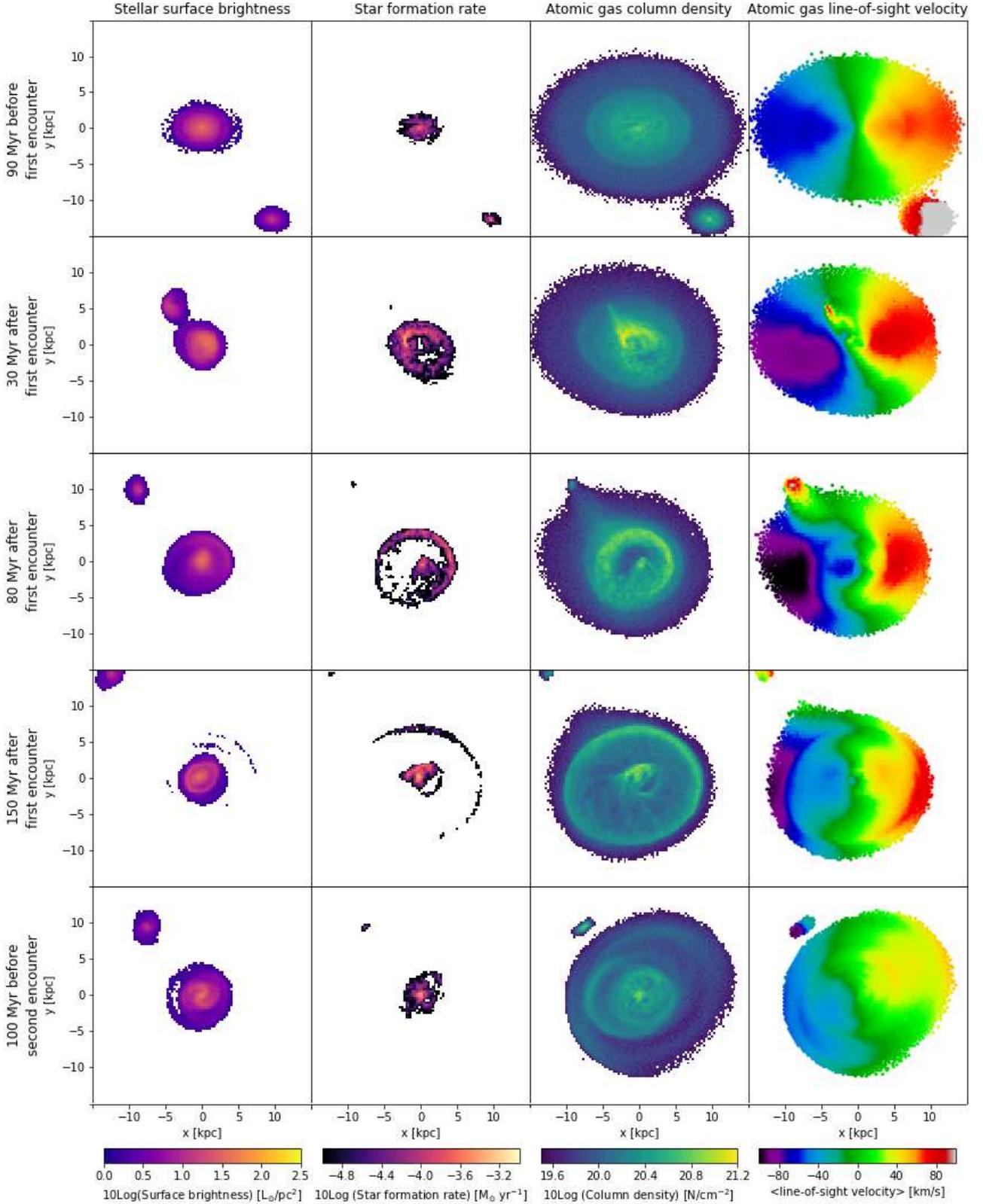,width=0.97\textwidth}
\end{center}
\caption{An example of a simulated head-on collision between a dwarf galaxy and a small companion, simultaneously reproducing the main observed  features  in  ESO 184$-$G82  (see Section \ref{sec:interaction} for details). The viewing angle for the simulation presented in this figure is 45 degrees from the plane of the primary dwarf galaxy. Each row corresponds to a snapshot in time of the dwarf--dwarf simulation, with time increasing from top to bottom. The four columns, from left to right, contains the distributions of, respectively, the stellar surface brightness, the SFR, the \hi column density, and the \hicm  velocity field.  The third row, corresponding to 80 Myr after the first collision, resembles  the observed properties of ESO 184$-$G82 and its companion, GALJ193510-524947. Note that in the first and last rows the secondary dwarf is moving towards the primary dwarf for a collision, and in the other three rows, it is moving away from the primary dwarf after a collision.    
}
\label{fig:sim}
\end{figure*}

Conversely, the collisional interaction simulations simultaneously reproduce all the above observational constraints.  Fig. \ref{fig:sim} presents the outcome of one of these simulations at five  different times; for each time sample, the different results are plotted in a single row. The four columns (from left to right) present,  respectively, the distributions of the stellar surface brightness, the SFR, the \hi column density, and the \hicm  velocity field.   
The viewing angle for the simulation presented in this figure is 45 degrees from the plane of the primary dwarf galaxy. 
The   primary and secondary dwarfs have   halo masses of $\rm 8 \times 10^8\,M_{\sun}$  and $\rm 3 \times 10^{10}\,M_{\sun}$ respectively. The dark matter particle masses for the primary and secondary dwarf are $\rm 2.2 \times 10^4\,M_{\sun}$ and $\rm 4.0 \times 10^4\,M_{\sun}$ while the  baryonic particle masses are  $\rm 1.6 \times 10^3\,M_{\sun}$ and $\rm 1.0 \times 10^3\,M_{\sun}$, respectively.

The top  row shows the   configuration of the system   in a snapshot corresponding to 90 Myr before the first collision, when  the secondary dwarf is at a distance of 20 kpc (projected distance of 16 kpc) from the primary galaxy.  The primary dwarf at this time has a stellar mass of $\rm \approx 7 \times 10^8\,M_{\sun}$  and a similar atomic gas mass.  The secondary dwarf is moving with a velocity of 100 \kms, on a direct collision course with the disk of the primary galaxy and at  an  angle of 70 degrees with respect to  the plane of the primary galaxy. 
The first collision occurs  at a point 2 kpc away from the  centre of the primary dwarf disk,   and   results in  the formation of annular shells  moving outwards from the collision point. The inner shells have larger velocities compared to the outer shells. The shock caused by this difference in velocities results in  the compression of gas and as a consequence, in the enhancement of  star formation.  The high density gas and the  enhancement of star formation  in the primary dwarf can be seen in the  second snapshot (the second row of Fig. \ref{fig:sim}) which shows the  system  30 Myr after the first collision.  By this time, star formation in the primary dwarf has  increased  by more than a factor of three compared to that before the collision.

\begin{figure}
\begin{center}
\psfig{file=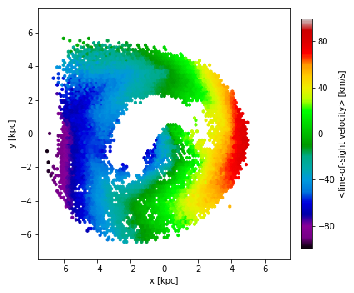,width=0.47\textwidth}
\end{center}
\vskip -7 mm
\caption{The simulated  velocity field of the gas ring in the primary dwarf 80 Myr after the first collision. 
This ring contains the atomic gas with column densities above $\rm 2 \times 10^{20}\,cm^{-2}$. Velocity gradients of a few tens of \kms\, can be seen to exist within sub-kpc scales, similar to what is observed in ESO 184$-$G8 (see the bottom-right panel of Fig. \ref{fig:ov}).      
}
\label{fig:ring-vel}
\end{figure}

With time, the  shock wave    moves  outwards  from the point of collision and forms  an asymmetric ring   of  dense atomic gas in the primary dwarf. This ring   is clearly visible  in the  third snapshot (the third row of Fig. \ref{fig:sim}) which   corresponds to  80 Myr after the first collision.  By this time the formation of molecular gas in the dense atomic gas ring has enhanced  the  star formation in the ring. At the same time, the lack of sufficient    inflow of gas to the centre of the primary dwarf has resulted in very little or  no  star formation enhancement in the centre of the galaxy. 
The secondary dwarf is at a distance of 17 kpc (projected distance of 13 kpc) from the primary galaxy and is moving away from it. The atomic gas which has been dragged out from the  primary galaxy is visible as the extension of gas between the two galaxies.  The overall   stellar  and gas distributions of the dwarf--dwarf system at this snapshot   matches those of ESO 184$-$G82 and  GALJ193510-524947. The velocity field of the atomic gas  too resembles the velocity map of the atomic gas in ESO 184$-$G82, presented in the bottom-middle panel of Fig. \ref{fig:ov}. The dense gas in the region of the ring   demonstrates velocity gradients of a few tens of \kms\,  over  sub-kpc scales. Such large velocity gradients  result in the formation of massive  gas clouds and hence large knots of star formation. These  massive gas clouds as well as large star formation knots  are resolved in our simulations and are visible  in the third row of   Fig. \ref{fig:sim}. The large velocity gradients can also be clearly seen in the  velocity distribution of the gas ring shown separately  in Fig. \ref{fig:ring-vel}. 
 
The radius of the gas ring continues to  increase with time and its density contrast decreases as can be seen in the fourth snapshot (the fourth row in Fig. \ref{fig:sim}) which corresponds to 150 Myr after the first collision. At this time the secondary galaxy is moving further away from the primary dwarf.   
The fifth  snapshot (the fifth row in Fig. \ref{fig:sim}), corresponding to 100 Myr before the second collision,  when the secondary galaxy, after reaching its apocentre,  is moving back towards the primary galaxy for the second collision. By this time the distribution of atomic gas in the primary galaxy has become more uniform, with no depression in the central regions. Also, the star formation activity is only present in the centre of the galaxy.

%-------------------------------------------------------------------------------------------

\section{Discussion}
\label{sec:dis}

The gas and star formation properties of ESO 184$-$G82, along with the identification of a companion galaxy, GALJ193510-524947, in its vicinity  provide clear evidence for interaction between the two galaxies. Our simulations support the paradigm in which  ESO 184$-$G82 had a head-on collision with its companion which led to the formation of the high column density gas ring observed in our \hi map.  
This  ring, with large velocity gradients (as high as 40 \kms\, over sub-kpc scales), is an ideal site for formation of  massive  Giant Molecular Clouds and hence   super star clusters (SSCs). This, consequently, increases the  probability of formation of  massive stars in the star forming regions within the gas ring and in turn increases the chance of a GRB progenitor to form. 
Note that this paradigm is for the formation of the GRB progenitor, and may not have any  bearing on the stellar explosion mechanism which ultimately produces the  GRB. 

A comparison between the stellar mass distributions of GRB hosts   and  the cosmic star formation rate density  shows  that most  GRBs occur in galaxies with stellar masses lower than those of galaxies  responsible for the bulk of star formation \citep[e.g. galaxies with $M_*=10^{10.0-10.7}M_{\odot}$ at $z<1$,][]{Conroy09-2009ApJ...696..620C}. This is especially the case for GRBs at $z \lesssim 2$ \citep[see][]{Perley16-2016ApJ...817....8P}. Hence, the processes that create these energetic explosions are not linked in a trivial way to the SFRs of their host galaxies. 
There must be other factors which drive the formation of GRB progenitors in galaxies. 

The typical low mass and metallicity of GRB hosts has been widely interpreted as low metallicity being the primary requirement for GRB formation. 
However, this has been called into question by the  detection of a large number of GRBs in massive, metal-rich galaxies \citep[][]{Savaglio12-2012MNRAS.420..627S, Svensson12-2012MNRAS.421...25S, Elliott13-2013A&A...556A..23E, Hunt14-2014A&A...565A.112H, Schady15-2015A&A...579A.126S, Perley13-2013ApJ...778..128P}.  
A less addressed, but more likely, factor is high SFR densities. 
\citet[][]{Kelly14-2014ApJ...789...23K} found the SFR  surface densities of GRB hosts to be higher than those of field galaxies, implying that they have high SFR densities. 
Regions with high star formation densities are expected to power strong gas outflows  with large velocities   \citep[][]{Lagos13-2013MNRAS.436.1787L, Sharma17-2017MNRAS.468.2176S}. The large galaxy outflow velocities deduced  from the  velocity widths of the interstellar medium  absorption lines in GRB hosts  \citep[detected in  GRB afterglows, see][and references therein]{Arabsalmani18-2018MNRAS.473.3312A} are therefore consistent with  the hypothesis of GRBs originating in regions with high SFR densities.

This picture is in agreement with   GRB formation models. 
There are two main models proposed for the generation of  relativistic jets in GRBs through core-collapse of massive stars. 
In one, a single progenitor  star with anomalously rapid rotation forms the GRB \citep[][]{MacFadyen99-1999ApJ...524..262M, Yoon05-2005A&A...443..643Y, Hirschi05-2005A&A...443..581H, Woosley06-2006ApJ...637..914W}, 
while in the other, the GRB is associated with the core collapse of a massive star  stripped by a companion in a close binary system \citep[][]{Usov92-1992Natur.357..472U, Izzard04-2004MNRAS.348.1215I, Podsiadlowski04-2004ApJ...607L..17P, Fryer05-2005ApJ...623..302F, Detmers08-2008A&A...484..831D, Podsiadlowski10-2010MNRAS.406..840P, Tout11-2011MNRAS.410.2458T, Kinugawa17-2017ApJ...849L..29K}. 
The single star model requires the progenitor to retain angular momentum (necessary for GRB formation) and, at the same time, lose substantial mass (to develop into a hydrogen- and helium-poor star).  While the presence of metals helps the mass loss (and the removal of the hydrogen/helium envelope), this also carries away angular momentum.  This  contradiction is avoided in chemically homogeneous evolved progenitors, which however require much lower metallicities  than is observed in GRB hosts.
On the other hand, the metallicity constraints on the progenitor stars when GRBs are produced in close binaries  are more relaxed;  although these models predict a higher probability of GRB formation in metal-poor progenitors, they do not 
rule out high (e.g. solar or super-solar) metallicities in the progenitors \citep[e.g., see][]{Podsiadlowski10-2010MNRAS.406..840P, Tout11-2011MNRAS.410.2458T}.        
It is notable that  \citet[][]{Sana14-2014ApJS..215...15S} found  massive stars to form nearly exclusively in multiple systems \citep[see also][]{Mason09-2009AJ....137.3358M, Sana12-2012Sci...337..444S}. Formation of GRB progenitors in massive and dense SSCs is particularly  in agreement with the models  in which  GRBs forms through dynamical interactions and collisions  of massive stars  in dense  environments  \citep[see for e.g.,][]{Heuvel13-2013ApJ...779..114V}.

The   link between high SFR densities  and  GRB progenitors is also supported by the observed top heavy initial mass function (IMF) in regions with high SFR densities. There are several lines of evidence in the local Universe  indicating that more massive stars are found in regions  with  high SFR densities than would be expected from a standard IMF \citep[e.g. Salpeter, Chabrier, etc;][]{Dabringhausen09-2009MNRAS.394.1529D, Dabringhausen12-2012ApJ...747...72D, Banerjee12-2012A&A...547A..23B, Marks12-2012MNRAS.422.2246M, Peacock17-2017ApJ...841...28P}. 
\citet[][]{Schneider17-2018Sci...359...69S} studied a compact and bright H{\sc ii} region in the  Large  Magellanic Cloud whose properties may closely replicate starbursts at high  redshifts. They  found it to contain $32\%$ more stars with masses larger than 30 $M_{\odot}$ than expected from a standard IMF. This is in line with the findings of \citet[][]{Zhang18-2018Natur.558..260Z} who recently investigated  four submillimetre galaxies at $z=2-3$ and found evidence for a top heavy IMF in  all of them.  
Therefore, regions with compact and intense   star formation, such as massive SSCs,  are the  likely  birth-place of GRB progenitors \citep[also see][]{Chen07-2007ApJ...668..384C}.

Massive SSCs    are known to be  common in  interacting systems  \citep[][]{Elmegreen-1993ApJ...412...90E, deGrijs03-2003NewA....8..155D, Bastian08-2008MNRAS.390..759B, Renaud-2018NewAR..81....1R}. The absence of gravitational shear and the increased turbulence  in interacting systems  are thought  to aid the gravitational collapse of massive amounts of gas into massive and compact GMCs, which may  subsequently form  SSCs  \citep[see][]{Elmegreen00-2000AJ....120..630E,  Weidner10-2010ApJ...724.1503W, Teyssier10-2010ApJ...720L.149T, Elmegreen17-2017ApJ...841...43E}.  
The large  velocity dispersion of the interstellar gas in interacting systems not only  increases  the Jeans mass  (which in turn   results  in the formation of  massive clumps), but also  heightens  the temperature of the clouds, thus  shifting the IMF toward more massive stars \citep[see][]{Elmegreen-1993ApJ...412...90E}. 
Collisional  encounters between   galaxies, though rare compared to  tidal interactions,  are   more efficient in triggering the  formation of massive and compact SSCs \citep[][]{Struck96-1996AJ....112.1868S, Burkert05-2005ApJ...628..231B, Elmegreen06-2006ApJ...651..676E, Pellerin10-2010AJ....139.1369P}.  In a recent study \citet[][]{Renaud18-2018MNRAS.473..585R} showed  that head-on collisions produce fewer, but larger  SSCs, compared to  tidal  interactions. 

The above  evidence suggests   a natural link between the host galaxies of GRBs and interacting systems. Such a link seems to be   especially likely at  $z \lesssim 1$ since interactions appear to play a dominant  role in  the formation of massive and compact SSCs at low redshifts.  At higher redshifts, the  high gas fractions of  galaxies  can cause gravitational instability leading  to  the     collapse of large amounts of gas into  massive and dense clumps. Recent studies however   indicate that, at high redshifts too,  violent mechanisms  such as  major or  minor mergers are required to generate  strong concentrations of gas \citep[see][and references therein]{Elbaz18-2018A&A...616A.110E}.

The  typical high sSFR values of  GRB host galaxies \citep[e.g.,][]{Sokolov01-2001A&A...372..438S, Chary02-2002ApJ...566..229C, Christensen04-2004A&A...425..913C, Savaglio09-2009ApJ...691..182S, Svensson10-2010MNRAS.405...57S,  Perley15-2015ApJ...801..102P}, suggesting    a recent boost in their star formation, are consistent  with the existence of a  link between GRB hosts and interacting systems. 
Interactions are known to  enhance the  star formation activities of galaxies and  increase  their  SFR up to an order of magnitude higher  \citep[e.g.,][]{Renaud14-2014MNRAS.442L..33R, Pan18-2018arXiv181010162P}. 
Earlier studies have found indications of interactions and mergers in GRB host galaxies, but  the evidence has not been unambiguous. 
\citet{Chary02-2002ApJ...566..229C} found  6 of the 11 GRB hosts in their sample to be disturbed or to have candidate companion galaxies.  Spectroscopic studies are required in order to, as the first step, confirm whether the candidate  companions  are indeed at the redshift of the GRB hosts. 
\citet[][]{Chen12-2012MNRAS.419.3039C} performed spectroscopic studies in the fields of two GRBs at $z=1.5$ and $z=2.6$ , and in both cases found  a few galaxies with small separations in projected distance and velocity space,  indicating them to belong to interacting systems.
\citet{Wainwright07-2007ApJ...657..367W}  reported the morphology of $30\%$ of the 42 hosts in their sample to show signs of interaction, with an additional $30\%$ exhibiting irregular and asymmetric structure.   \citet[][]{Savaglio09-2009ApJ...691..182S} found the morphology of 10 out of  22 GRB hosts to be asymmetric or similar to merger remnants  \citep[see also][and references therein]{Savaglio15-2015JHEAp...7...95S}. However,  interpreting the morphological signatures of interaction can be challenging, especially  at high redshifts. 
Absorption studies  have also indicated possible ongoing interactions in GRB hosts.  
\citet[][]{Savaglio12-2012MNRAS.420..627S} reported the presence of strong double absorption systems with small velocity  separation in 5 out of 40 GRB spectra, compared to 18 cases out of 500 for absorbers in sightlines towards  quasars \citep[see also][]{Wiseman17-2017A&A...607A.107W, Arabsalmani18-2018MNRAS.473.3312A}. Emission studies  are required to confirm whether the  multi-component systems detected in the pencil-beams of GRB afterglows are associated with interacting systems or 
if they are related to other phenomena  such as strong outflowing gas in the host galaxies \citep[see][]{Arabsalmani18-2018MNRAS.tmp..190A, Arabsalmani18-2018MNRAS.473.3312A}.

In this study we  find clear evidence for an ongoing interaction between the host galaxy of GRB 980425 and its  companion, through a detailed \hicm study of the distribution and kinematics of atomic gas in the GRB host galaxy. In addition, our simulations show that the gas and star formation properties of the host galaxy of GRB 980425 can be reproduced by a collisional interaction with its companion galaxy.  
The similar ages of massive stars in the actively star forming regions of the host galaxy of GRB 980425 and its companion galaxy  suggest that the interaction between the two galaxies  has triggered the recent star formation in them. This is a clear case linking  interaction driven star formation to a GRB event.

\section{summary}
\label{sec:sum}
We have used the GMRT to map the \hicm emission from ESO 184$-$G82, the $z=0.0087$ host galaxy of GRB 980425/SN1998bw. The \hicm intensity images and velocity distribution yield clear evidence that ESO 184$-$G82 is undergoing an interaction with a companion galaxy: these include the detection of an \hi knot to the North-East of ESO 184-G82, an extended \hi structure extending from ESO 184-G82 towards the \hi knot, the disturbed \hi velocity field, and finally the presence of a high column density \hi ring, likely a collisional ring, around the optical centre of ESO 184$-$G82, passing through the actively star-forming regions of the galaxy. 
We use VLT/FORS, HST, and Spitzer optical and NIR imaging to identify a small galaxy coincident with the \hi knot detected in the GMRT \hicm image, at a projected distance of 13 kpc from the centre of the GRB host galaxy. We find the companion galaxy to be a gas-rich star-forming dwarf galaxy, with an SFR of  $0.004$ $\rm M_{\odot}\, yr^{-1}$, a gas mass of $\rm 10^{7.1 \pm 0.1}\,M_{\odot}$, and  a stellar mass of $\rm 10^{7.0 \pm 0.3}\,M_{\odot}$. 

Head-on collisions produce star-forming gaseous rings with high surface densities and velocity dispersions, leading to the formation of  massive SSCs  in the ring. At the same time they cause little (or no) star formation enhancement in the centre of the galaxies, unlike classical tidal interactions. Our  simulations of dwarf-dwarf mergers illustrate this process and show that a head-on collision  can reproduce the main observed gas and star formation features  of ESO 184$-$G82 simultaneously. This is  while   it is difficult to explain  the  observed properties such as the presence of the gas ring and  the absence  of the star formation enhancement in the centre  of ESO 184$-$G82  via a tidal encounter. 
Our findings therefore suggest that the collision between ESO 184$-$G82 and its companion galaxy has lead to the formation of dense and massive  SSCs in which the GRB progenitor must have formed.

\section*{Acknowledgments}
M.A. and S.R.  would like to thank Bruce Elmegreen,  Francoise Combes, Diane Cormier, and David Elbaz for  valuable discussions. 
We thank the staff of the GMRT for making  these observations possible. The GMRT is run by the National Centre for Radio Astrophysics  of  the  Tata  Institute  of  Fundamental  Research. We  acknowledge using  data based on observations collected at the European Southern Observatory under ESO programs 064.H-0375(A), 066.D-0576(A), and 165.H-0464(A).  
This work is based in part on observations made with the Spitzer Space Telescope, which is operated by the Jet Propulsion Laboratory, California Institute of Technology under a contract with NASA, and also  data made with the NASA/ESA Hubble Space Telescope, obtained from the data archive at the Space Telescope Science Institute. 
M.A. acknowledges support from  UnivEarthS Labex program at Sorbonne Paris Cit\'e (ANR-10-LABX-0023 and ANR-11-IDEX-0005-02). 
N.K. acknowledges support from the Department of Science and Technology via  a Swarnajayanti Fellowship (DST/SJF/PSA-01/2012-13). L.C. is supported by DFF -- 4090-00079. 
The Flatiron Institute is supported by the Simons Foundation. The Cosmic Dawn Center is funded by the DNRF. STScI is operated by the Association of Universities for Research in Astronomy, Inc. under NASA contract NAS 5-26555.

\bibliographystyle{mnras}
\bibliography{adssample}

\begin{thebibliography}{}
\makeatletter
\relax
\def\mn@urlcharsother{\let\do\@makeother \do\$\do\&\do\#\do\^\do\_\do\%\do\~}
\def\mn@doi{\begingroup\mn@urlcharsother \@ifnextchar [ {\mn@doi@}
  {\mn@doi@[]}}
\def\mn@doi@[#1]#2{\def\@tempa{#1}\ifx\@tempa\@empty \href
  {http://dx.doi.org/#2} {doi:#2}\else \href {http://dx.doi.org/#2} {#1}\fi
  \endgroup}
\def\mn@eprint#1#2{\mn@eprint@#1:#2::\@nil}
\def\mn@eprint@arXiv#1{\href {http://arxiv.org/abs/#1} {{\tt arXiv:#1}}}
\def\mn@eprint@dblp#1{\href {http://dblp.uni-trier.de/rec/bibtex/#1.xml}
  {dblp:#1}}
\def\mn@eprint@#1:#2:#3:#4\@nil{\def\@tempa {#1}\def\@tempb {#2}\def\@tempc
  {#3}\ifx \@tempc \@empty \let \@tempc \@tempb \let \@tempb \@tempa \fi \ifx
  \@tempb \@empty \def\@tempb {arXiv}\fi \@ifundefined
  {mn@eprint@\@tempb}{\@tempb:\@tempc}{\expandafter \expandafter \csname
  mn@eprint@\@tempb\endcsname \expandafter{\@tempc}}}

\bibitem[\protect\citeauthoryear{{Aloy}, {M{\"u}ller}, {Ib{\'a}{\~n}ez},
  {Mart{\'{\i}}}  \& {MacFadyen}}{{Aloy}
  et~al.}{2000}]{Aloy00-2000ApJ...531L.119A}
{Aloy} M.~A.,  {M{\"u}ller} E.,  {Ib{\'a}{\~n}ez} J.~M.,  {Mart{\'{\i}}} J.~M.,
    {MacFadyen} A.,  2000, \mn@doi [\apjl] {10.1086/312537}, \href
  {http://adsabs.harvard.edu/abs/2000ApJ...531L.119A} {531, L119}

\bibitem[\protect\citeauthoryear{{Appleton} \& {Struck-Marcell}}{{Appleton} \&
  {Struck-Marcell}}{1996}]{Appleton96-1996FCPh...16..111A}
{Appleton} P.~N.,  {Struck-Marcell} C.,  1996, \fcp, \href
  {http://adsabs.harvard.edu/abs/1996FCPh...16..111A} {16, 111}

\bibitem[\protect\citeauthoryear{{Arabsalmani}, {Roychowdhury}, {Zwaan},
  {Kanekar}  \& {Micha{\l}owski}}{{Arabsalmani}
  et~al.}{2015}]{Arabsalmani15-2015MNRAS.454L..51A}
{Arabsalmani} M.,  {Roychowdhury} S.,  {Zwaan} M.~A.,  {Kanekar} N.,
  {Micha{\l}owski} M.~J.,  2015, \mn@doi [\mnras] {10.1093/mnrasl/slv118},
  \href {http://adsabs.harvard.edu/abs/2015MNRAS.454L..51A} {454, L51}

\bibitem[\protect\citeauthoryear{{Arabsalmani} et~al.,}{{Arabsalmani}
  et~al.}{2018a}]{Arabsalmani18-2018MNRAS.tmp..190A}
{Arabsalmani} M.,  et~al., 2018a, \mn@doi [\mnras] {10.1093/mnras/sty194},
  \href {http://adsabs.harvard.edu/abs/2018MNRAS.tmp..190A} {}

\bibitem[\protect\citeauthoryear{{Arabsalmani} et~al.,}{{Arabsalmani}
  et~al.}{2018b}]{Arabsalmani18-2018MNRAS.473.3312A}
{Arabsalmani} M.,  et~al., 2018b, \mn@doi [\mnras] {10.1093/mnras/stx2451},
  \href {http://adsabs.harvard.edu/abs/2018MNRAS.473.3312A} {473, 3312}

\bibitem[\protect\citeauthoryear{{Arnouts}, {Cristiani}, {Moscardini},
  {Matarrese}, {Lucchin}, {Fontana}  \& {Giallongo}}{{Arnouts}
  et~al.}{1999}]{1999MNRAS.310..540A}
{Arnouts} S.,  {Cristiani} S.,  {Moscardini} L.,  {Matarrese} S.,  {Lucchin}
  F.,  {Fontana} A.,   {Giallongo} E.,  1999, \mn@doi [\mnras]
  {10.1046/j.1365-8711.1999.02978.x}, \href
  {http://adsabs.harvard.edu/abs/1999MNRAS.310..540A} {310, 540}

\bibitem[\protect\citeauthoryear{{Banerjee} \& {Kroupa}}{{Banerjee} \&
  {Kroupa}}{2012}]{Banerjee12-2012A&A...547A..23B}
{Banerjee} S.,  {Kroupa} P.,  2012, \mn@doi [\aap]
  {10.1051/0004-6361/201218972}, \href
  {http://adsabs.harvard.edu/abs/2012A%26A...547A..23B} {547, A23}

\bibitem[\protect\citeauthoryear{{Barnes} \& {Hernquist}}{{Barnes} \&
  {Hernquist}}{1991}]{Barnes91-1991ApJ...370L..65B}
{Barnes} J.~E.,  {Hernquist} L.~E.,  1991, \mn@doi [\apjl] {10.1086/185978},
  \href {http://adsabs.harvard.edu/abs/1991ApJ...370L..65B} {370, L65}

\bibitem[\protect\citeauthoryear{{Bastian}}{{Bastian}}{2008}]{Bastian08-2008MNRAS.390..759B}
{Bastian} N.,  2008, \mn@doi [\mnras] {10.1111/j.1365-2966.2008.13775.x}, \href
  {http://adsabs.harvard.edu/abs/2008MNRAS.390..759B} {390, 759}

\bibitem[\protect\citeauthoryear{{Bloom}, {Kulkarni}  \& {Djorgovski}}{{Bloom}
  et~al.}{2002}]{Bloom02-2002AJ....123.1111B}
{Bloom} J.~S.,  {Kulkarni} S.~R.,   {Djorgovski} S.~G.,  2002, \mn@doi [\aj]
  {10.1086/338893}, \href {http://adsabs.harvard.edu/abs/2002AJ....123.1111B}
  {123, 1111}

\bibitem[\protect\citeauthoryear{{Bournaud} \& {Combes}}{{Bournaud} \&
  {Combes}}{2003}]{Bournaud03-2003A&A...401..817B}
{Bournaud} F.,  {Combes} F.,  2003, \mn@doi [\aap]
  {10.1051/0004-6361:20030150}, \href
  {http://cdsads.u-strasbg.fr/abs/2003A%26A...401..817B} {401, 817}

\bibitem[\protect\citeauthoryear{{Brinchmann}, {Charlot}, {White}, {Tremonti},
  {Kauffmann}, {Heckman}  \& {Brinkmann}}{{Brinchmann}
  et~al.}{2004}]{Brinchmann04-2004MNRAS.351.1151B}
{Brinchmann} J.,  {Charlot} S.,  {White} S.~D.~M.,  {Tremonti} C.,  {Kauffmann}
  G.,  {Heckman} T.,   {Brinkmann} J.,  2004, \mn@doi [\mnras]
  {10.1111/j.1365-2966.2004.07881.x}, \href
  {http://adsabs.harvard.edu/abs/2004MNRAS.351.1151B} {351, 1151}

\bibitem[\protect\citeauthoryear{{Bruzual} \& {Charlot}}{{Bruzual} \&
  {Charlot}}{2003}]{Bruzual03-2003MNRAS.344.1000B}
{Bruzual} G.,  {Charlot} S.,  2003, \mn@doi [\mnras]
  {10.1046/j.1365-8711.2003.06897.x}, \href
  {http://adsabs.harvard.edu/abs/2003MNRAS.344.1000B} {344, 1000}

\bibitem[\protect\citeauthoryear{{Burkert}, {Brodie}  \& {Larsen}}{{Burkert}
  et~al.}{2005}]{Burkert05-2005ApJ...628..231B}
{Burkert} A.,  {Brodie} J.,   {Larsen} S.,  2005, \mn@doi [\apj]
  {10.1086/430698}, \href {http://cdsads.u-strasbg.fr/abs/2005ApJ...628..231B}
  {628, 231}

\bibitem[\protect\citeauthoryear{{Buta}}{{Buta}}{1999}]{Buta99-1999Ap&SS.269...79B}
{Buta} R.,  1999, \mn@doi [\apss] {10.1023/A:1017059621651}, \href
  {http://adsabs.harvard.edu/abs/1999Ap%26SS.269...79B} {269, 79}

\bibitem[\protect\citeauthoryear{{Buta} \& {Combes}}{{Buta} \&
  {Combes}}{1996}]{Buta96-1996FCPh...17...95B}
{Buta} R.,  {Combes} F.,  1996, \fcp, \href
  {http://adsabs.harvard.edu/abs/1996FCPh...17...95B} {17, 95}

\bibitem[\protect\citeauthoryear{{Castro Cer{\'o}n}, {Micha{\l}owski},
  {Hjorth}, {Malesani}, {Gorosabel}, {Watson}, {Fynbo}  \& {Morales
  Calder{\'o}n}}{{Castro Cer{\'o}n}
  et~al.}{2010}]{Castroceron10-2010ApJ...721.1919C}
{Castro Cer{\'o}n} J.~M.,  {Micha{\l}owski} M.~J.,  {Hjorth} J.,  {Malesani}
  D.,  {Gorosabel} J.,  {Watson} D.,  {Fynbo} J.~P.~U.,   {Morales
  Calder{\'o}n} M.,  2010, \mn@doi [\apj] {10.1088/0004-637X/721/2/1919}, \href
  {http://adsabs.harvard.edu/abs/2010ApJ...721.1919C} {721, 1919}

\bibitem[\protect\citeauthoryear{{Chabrier}}{{Chabrier}}{2003}]{2003PASP..115..763C}
{Chabrier} G.,  2003, \mn@doi [\pasp] {10.1086/376392}, \href
  {http://adsabs.harvard.edu/abs/2003PASP..115..763C} {115, 763}

\bibitem[\protect\citeauthoryear{{Chary}, {Becklin}  \& {Armus}}{{Chary}
  et~al.}{2002}]{Chary02-2002ApJ...566..229C}
{Chary} R.,  {Becklin} E.~E.,   {Armus} L.,  2002, \mn@doi [\apj]
  {10.1086/337964}, \href {http://adsabs.harvard.edu/abs/2002ApJ...566..229C}
  {566, 229}

\bibitem[\protect\citeauthoryear{{Chen}}{{Chen}}{2012}]{Chen12-2012MNRAS.419.3039C}
{Chen} H.-W.,  2012, \mn@doi [\mnras] {10.1111/j.1365-2966.2011.19944.x}, \href
  {http://adsabs.harvard.edu/abs/2012MNRAS.419.3039C} {419, 3039}

\bibitem[\protect\citeauthoryear{{Chen}, {Prochaska}  \& {Bloom}}{{Chen}
  et~al.}{2007}]{Chen07-2007ApJ...668..384C}
{Chen} H.-W.,  {Prochaska} J.~X.,   {Bloom} J.~S.,  2007, \mn@doi [\apj]
  {10.1086/521021}, \href {http://adsabs.harvard.edu/abs/2007ApJ...668..384C}
  {668, 384}

\bibitem[\protect\citeauthoryear{{Christensen}, {Hjorth}  \&
  {Gorosabel}}{{Christensen} et~al.}{2004}]{Christensen04-2004A&A...425..913C}
{Christensen} L.,  {Hjorth} J.,   {Gorosabel} J.,  2004, \mn@doi [\aap]
  {10.1051/0004-6361:20040361}, \href
  {http://adsabs.harvard.edu/abs/2004A%26A...425..913C} {425, 913}

\bibitem[\protect\citeauthoryear{{Christensen}, {Vreeswijk}, {Sollerman},
  {Th{\"o}ne}, {Le Floc'h}  \& {Wiersema}}{{Christensen}
  et~al.}{2008}]{Christensen08-2008A&A...490...45C}
{Christensen} L.,  {Vreeswijk} P.~M.,  {Sollerman} J.,  {Th{\"o}ne} C.~C.,  {Le
  Floc'h} E.,   {Wiersema} K.,  2008, \mn@doi [\aap]
  {10.1051/0004-6361:200809896}, \href
  {http://adsabs.harvard.edu/abs/2008A%26A...490...45C} {490, 45}

\bibitem[\protect\citeauthoryear{{Conroy} \& {Wechsler}}{{Conroy} \&
  {Wechsler}}{2009}]{Conroy09-2009ApJ...696..620C}
{Conroy} C.,  {Wechsler} R.~H.,  2009, \mn@doi [\apj]
  {10.1088/0004-637X/696/1/620}, \href
  {http://adsabs.harvard.edu/abs/2009ApJ...696..620C} {696, 620}

\bibitem[\protect\citeauthoryear{{Cucchiara}, {Fumagalli}, {Rafelski},
  {Kocevski}, {Prochaska}, {Cooke}  \& {Becker}}{{Cucchiara}
  et~al.}{2015}]{Cucchiara15-2015ApJ...804...51C}
{Cucchiara} A.,  {Fumagalli} M.,  {Rafelski} M.,  {Kocevski} D.,  {Prochaska}
  J.~X.,  {Cooke} R.~J.,   {Becker} G.~D.,  2015, \mn@doi [\apj]
  {10.1088/0004-637X/804/1/51}, \href
  {http://adsabs.harvard.edu/abs/2015ApJ...804...51C} {804, 51}

\bibitem[\protect\citeauthoryear{{Dabringhausen}, {Kroupa}  \&
  {Baumgardt}}{{Dabringhausen}
  et~al.}{2009}]{Dabringhausen09-2009MNRAS.394.1529D}
{Dabringhausen} J.,  {Kroupa} P.,   {Baumgardt} H.,  2009, \mn@doi [\mnras]
  {10.1111/j.1365-2966.2009.14425.x}, \href
  {http://adsabs.harvard.edu/abs/2009MNRAS.394.1529D} {394, 1529}

\bibitem[\protect\citeauthoryear{{Dabringhausen}, {Kroupa}, {Pflamm-Altenburg}
  \& {Mieske}}{{Dabringhausen}
  et~al.}{2012}]{Dabringhausen12-2012ApJ...747...72D}
{Dabringhausen} J.,  {Kroupa} P.,  {Pflamm-Altenburg} J.,   {Mieske} S.,  2012,
  \mn@doi [\apj] {10.1088/0004-637X/747/1/72}, \href
  {http://adsabs.harvard.edu/abs/2012ApJ...747...72D} {747, 72}

\bibitem[\protect\citeauthoryear{{Dalla Vecchia} \& {Schaye}}{{Dalla Vecchia}
  \& {Schaye}}{2008}]{DallaVecchia08-2008MNRAS.387.1431D}
{Dalla Vecchia} C.,  {Schaye} J.,  2008, \mn@doi [\mnras]
  {10.1111/j.1365-2966.2008.13322.x}, \href
  {http://adsabs.harvard.edu/abs/2008MNRAS.387.1431D} {387, 1431}

\bibitem[\protect\citeauthoryear{{Detmers}, {Langer}, {Podsiadlowski}  \&
  {Izzard}}{{Detmers} et~al.}{2008}]{Detmers08-2008A&A...484..831D}
{Detmers} R.~G.,  {Langer} N.,  {Podsiadlowski} P.,   {Izzard} R.~G.,  2008,
  \mn@doi [\aap] {10.1051/0004-6361:200809371}, \href
  {http://adsabs.harvard.edu/abs/2008A%26A...484..831D} {484, 831}

\bibitem[\protect\citeauthoryear{{Di Matteo}, {Combes}, {Melchior}  \&
  {Semelin}}{{Di Matteo} et~al.}{2007}]{DiMatteo07-2007A&A...468...61D}
{Di Matteo} P.,  {Combes} F.,  {Melchior} A.-L.,   {Semelin} B.,  2007, \mn@doi
  [\aap] {10.1051/0004-6361:20066959}, \href
  {http://cdsads.u-strasbg.fr/abs/2007A%26A...468...61D} {468, 61}

\bibitem[\protect\citeauthoryear{{Di Matteo}, {Bournaud}, {Martig}, {Combes},
  {Melchior}  \& {Semelin}}{{Di Matteo}
  et~al.}{2008}]{DiMatteo08-2008A&A...492...31D}
{Di Matteo} P.,  {Bournaud} F.,  {Martig} M.,  {Combes} F.,  {Melchior} A.-L.,
   {Semelin} B.,  2008, \mn@doi [\aap] {10.1051/0004-6361:200809480}, \href
  {http://cdsads.u-strasbg.fr/abs/2008A%26A...492...31D} {492, 31}

\bibitem[\protect\citeauthoryear{{Elbaz} et~al.,}{{Elbaz}
  et~al.}{2018}]{Elbaz18-2018A&A...616A.110E}
{Elbaz} D.,  et~al., 2018, \mn@doi [\aap] {10.1051/0004-6361/201732370}, \href
  {http://adsabs.harvard.edu/abs/2018A%26A...616A.110E} {616, A110}

\bibitem[\protect\citeauthoryear{{Elliott} et~al.,}{{Elliott}
  et~al.}{2013}]{Elliott13-2013A&A...556A..23E}
{Elliott} J.,  et~al., 2013, \mn@doi [\aap] {10.1051/0004-6361/201220968},
  \href {http://adsabs.harvard.edu/abs/2013A%26A...556A..23E} {556, A23}

\bibitem[\protect\citeauthoryear{{Elmegreen} \& {Elmegreen}}{{Elmegreen} \&
  {Elmegreen}}{2006}]{Elmegreen06-2006ApJ...651..676E}
{Elmegreen} D.~M.,  {Elmegreen} B.~G.,  2006, \mn@doi [\apj] {10.1086/507863},
  \href {http://adsabs.harvard.edu/abs/2006ApJ...651..676E} {651, 676}

\bibitem[\protect\citeauthoryear{{Elmegreen}, {Kaufman}  \&
  {Thomasson}}{{Elmegreen} et~al.}{1993}]{Elmegreen-1993ApJ...412...90E}
{Elmegreen} B.~G.,  {Kaufman} M.,   {Thomasson} M.,  1993, \mn@doi [\apj]
  {10.1086/172903}, \href {http://adsabs.harvard.edu/abs/1993ApJ...412...90E}
  {412, 90}

\bibitem[\protect\citeauthoryear{{Elmegreen} et~al.,}{{Elmegreen}
  et~al.}{2000}]{Elmegreen00-2000AJ....120..630E}
{Elmegreen} B.~G.,  et~al., 2000, \mn@doi [\aj] {10.1086/301462}, \href
  {http://adsabs.harvard.edu/abs/2000AJ....120..630E} {120, 630}

\bibitem[\protect\citeauthoryear{{Elmegreen}, {Elmegreen}, {Kaufman}, {Brinks},
  {Struck}, {Bournaud}, {Sheth}  \& {Juneau}}{{Elmegreen}
  et~al.}{2017}]{Elmegreen17-2017ApJ...841...43E}
{Elmegreen} D.~M.,  {Elmegreen} B.~G.,  {Kaufman} M.,  {Brinks} E.,  {Struck}
  C.,  {Bournaud} F.,  {Sheth} K.,   {Juneau} S.,  2017, \mn@doi [\apj]
  {10.3847/1538-4357/aa6ba5}, \href
  {http://cdsads.u-strasbg.fr/abs/2017ApJ...841...43E} {841, 43}

\bibitem[\protect\citeauthoryear{{Foley}, {Watson}, {Gorosabel}, {Fynbo},
  {Sollerman}, {McGlynn}, {McBreen}  \& {Hjorth}}{{Foley}
  et~al.}{2006}]{Foley06-2006A&A...447..891F}
{Foley} S.,  {Watson} D.,  {Gorosabel} J.,  {Fynbo} J.~P.~U.,  {Sollerman} J.,
  {McGlynn} S.,  {McBreen} B.,   {Hjorth} J.,  2006, \mn@doi [\aap]
  {10.1051/0004-6361:20054382}, \href
  {http://adsabs.harvard.edu/abs/2006A%26A...447..891F} {447, 891}

\bibitem[\protect\citeauthoryear{{Fruchter} et~al.,}{{Fruchter}
  et~al.}{2006}]{Fruchter06-2006Natur.441..463F}
{Fruchter} A.~S.,  et~al., 2006, \mn@doi [\nat] {10.1038/nature04787}, \href
  {http://adsabs.harvard.edu/abs/2006Natur.441..463F} {441, 463}

\bibitem[\protect\citeauthoryear{{Fryer} \& {Heger}}{{Fryer} \&
  {Heger}}{2005}]{Fryer05-2005ApJ...623..302F}
{Fryer} C.~L.,  {Heger} A.,  2005, \mn@doi [\apj] {10.1086/428379}, \href
  {http://adsabs.harvard.edu/abs/2005ApJ...623..302F} {623, 302}

\bibitem[\protect\citeauthoryear{{Fynbo} et~al.,}{{Fynbo}
  et~al.}{2000}]{Fynbo00}
{Fynbo} J.~U.,  et~al., 2000, \mn@doi [\apjl] {10.1086/312942}, \href
  {http://adsabs.harvard.edu/abs/2000ApJ...542L..89F} {542, L89}

\bibitem[\protect\citeauthoryear{{Fynbo} et~al.,}{{Fynbo}
  et~al.}{2006a}]{Fynbo06-2006Natur.444.1047F}
{Fynbo} J.~P.~U.,  et~al., 2006a, \mn@doi [\nat] {10.1038/nature05375}, \href
  {http://adsabs.harvard.edu/abs/2006Natur.444.1047F} {444, 1047}

\bibitem[\protect\citeauthoryear{{Fynbo} et~al.,}{{Fynbo}
  et~al.}{2006b}]{Fynbo06-2006A&A...451L..47F}
{Fynbo} J.~P.~U.,  et~al., 2006b, \mn@doi [\aap] {10.1051/0004-6361:20065056},
  \href {http://adsabs.harvard.edu/abs/2006A%26A...451L..47F} {451, L47}

\bibitem[\protect\citeauthoryear{{Galama} et~al.,}{{Galama}
  et~al.}{1998}]{Galama98-1998Natur.395..670G}
{Galama} T.~J.,  et~al., 1998, \mn@doi [\nat] {10.1038/27150}, \href
  {http://adsabs.harvard.edu/abs/1998Natur.395..670G} {395, 670}

\bibitem[\protect\citeauthoryear{{Genel} et~al.,}{{Genel}
  et~al.}{2014}]{Genel14-2014MNRAS.445..175G}
{Genel} S.,  et~al., 2014, \mn@doi [\mnras] {10.1093/mnras/stu1654}, \href
  {http://adsabs.harvard.edu/abs/2014MNRAS.445..175G} {445, 175}

\bibitem[\protect\citeauthoryear{{Graham} \& {Fruchter}}{{Graham} \&
  {Fruchter}}{2013}]{Graham13-2013ApJ...774..119G}
{Graham} J.~F.,  {Fruchter} A.~S.,  2013, \mn@doi [\apj]
  {10.1088/0004-637X/774/2/119}, \href
  {http://adsabs.harvard.edu/abs/2013ApJ...774..119G} {774, 119}

\bibitem[\protect\citeauthoryear{{Greisen}}{{Greisen}}{2003}]{Greisen03-2003ASSL..285..109G}
{Greisen} E.~W.,  2003, in {Heck} A.,  ed.,  Astrophysics and Space Science
  Library Vol. 285, Information Handling in Astronomy - Historical Vistas.
  p.~109, \mn@doi{10.1007/0-306-48080-8_7}

\bibitem[\protect\citeauthoryear{{Hammer}, {Flores}, {Schaerer},
  {Dessauges-Zavadsky}, {Le Floc'h}  \& {Puech}}{{Hammer}
  et~al.}{2006}]{Hammer06-2006A&A...454..103H}
{Hammer} F.,  {Flores} H.,  {Schaerer} D.,  {Dessauges-Zavadsky} M.,  {Le
  Floc'h} E.,   {Puech} M.,  2006, \mn@doi [\aap] {10.1051/0004-6361:20064823},
  \href {http://adsabs.harvard.edu/abs/2006A%26A...454..103H} {454, 103}

\bibitem[\protect\citeauthoryear{{Hibbard} \& {van Gorkom}}{{Hibbard} \& {van
  Gorkom}}{1996}]{Hibbard96-1996AJ....111..655H}
{Hibbard} J.~E.,  {van Gorkom} J.~H.,  1996, \mn@doi [\aj] {10.1086/117815},
  \href {http://adsabs.harvard.edu/abs/1996AJ....111..655H} {111, 655}

\bibitem[\protect\citeauthoryear{{Hirschi}, {Meynet}  \& {Maeder}}{{Hirschi}
  et~al.}{2005}]{Hirschi05-2005A&A...443..581H}
{Hirschi} R.,  {Meynet} G.,   {Maeder} A.,  2005, \mn@doi [\aap]
  {10.1051/0004-6361:20053329}, \href
  {http://adsabs.harvard.edu/abs/2005A%26A...443..581H} {443, 581}

\bibitem[\protect\citeauthoryear{{Hjorth} et~al.,}{{Hjorth}
  et~al.}{2003}]{Hjorth03-2003Natur.423..847H}
{Hjorth} J.,  et~al., 2003, \mn@doi [\nat] {10.1038/nature01750}, \href
  {http://adsabs.harvard.edu/abs/2003Natur.423..847H} {423, 847}

\bibitem[\protect\citeauthoryear{{Horellou} \& {Combes}}{{Horellou} \&
  {Combes}}{2001}]{Horellou01-2001Ap&SS.276.1141H}
{Horellou} C.,  {Combes} F.,  2001, \mn@doi [\apss] {10.1023/A:1017524632342},
  \href {http://adsabs.harvard.edu/abs/2001Ap%26SS.276.1141H} {276, 1141}

\bibitem[\protect\citeauthoryear{{Houck} et~al.,}{{Houck}
  et~al.}{2004}]{Houck04-2004ApJS..154...18H}
{Houck} J.~R.,  et~al., 2004, \mn@doi [\apjs] {10.1086/423134}, \href
  {http://adsabs.harvard.edu/abs/2004ApJS..154...18H} {154, 18}

\bibitem[\protect\citeauthoryear{{Hunt} et~al.,}{{Hunt}
  et~al.}{2014}]{Hunt14-2014A&A...565A.112H}
{Hunt} L.~K.,  et~al., 2014, \mn@doi [\aap] {10.1051/0004-6361/201323340},
  \href {http://adsabs.harvard.edu/abs/2014A%26A...565A.112H} {565, A112}

\bibitem[\protect\citeauthoryear{{Izzard}, {Ramirez-Ruiz}  \& {Tout}}{{Izzard}
  et~al.}{2004}]{Izzard04-2004MNRAS.348.1215I}
{Izzard} R.~G.,  {Ramirez-Ruiz} E.,   {Tout} C.~A.,  2004, \mn@doi [\mnras]
  {10.1111/j.1365-2966.2004.07436.x}, \href
  {http://adsabs.harvard.edu/abs/2004MNRAS.348.1215I} {348, 1215}

\bibitem[\protect\citeauthoryear{{Kelly}, {Filippenko}, {Modjaz}  \&
  {Kocevski}}{{Kelly} et~al.}{2014}]{Kelly14-2014ApJ...789...23K}
{Kelly} P.~L.,  {Filippenko} A.~V.,  {Modjaz} M.,   {Kocevski} D.,  2014,
  \mn@doi [\apj] {10.1088/0004-637X/789/1/23}, \href
  {http://adsabs.harvard.edu/abs/2014ApJ...789...23K} {789, 23}

\bibitem[\protect\citeauthoryear{{Kennicutt}}{{Kennicutt}}{1998}]{Kennicutt98-1998ApJ...498..541K}
{Kennicutt} Jr. R.~C.,  1998, \mn@doi [\apj] {10.1086/305588}, \href
  {http://adsabs.harvard.edu/abs/1998ApJ...498..541K} {498, 541}

\bibitem[\protect\citeauthoryear{{Kinugawa} \& {Asano}}{{Kinugawa} \&
  {Asano}}{2017}]{Kinugawa17-2017ApJ...849L..29K}
{Kinugawa} T.,  {Asano} K.,  2017, \mn@doi [\apjl] {10.3847/2041-8213/aa95bb},
  \href {http://adsabs.harvard.edu/abs/2017ApJ...849L..29K} {849, L29}

\bibitem[\protect\citeauthoryear{{Kr{\"u}hler} et~al.,}{{Kr{\"u}hler}
  et~al.}{2015}]{Kruhler15-2015A&A...581A.125K}
{Kr{\"u}hler} T.,  et~al., 2015, \mn@doi [\aap] {10.1051/0004-6361/201425561},
  \href {http://adsabs.harvard.edu/abs/2015A%26A...581A.125K} {581, A125}

\bibitem[\protect\citeauthoryear{{Kr{\"u}hler}, {Kuncarayakti}, {Schady},
  {Anderson}, {Galbany}  \& {Gensior}}{{Kr{\"u}hler}
  et~al.}{2017}]{Kruhler17-2017A&A...602A..85K}
{Kr{\"u}hler} T.,  {Kuncarayakti} H.,  {Schady} P.,  {Anderson} J.~P.,
  {Galbany} L.,   {Gensior} J.,  2017, \mn@doi [\aap]
  {10.1051/0004-6361/201630268}, \href
  {http://adsabs.harvard.edu/abs/2017A%26A...602A..85K} {602, A85}

\bibitem[\protect\citeauthoryear{{Lagos}, {Lacey}  \& {Baugh}}{{Lagos}
  et~al.}{2013}]{Lagos13-2013MNRAS.436.1787L}
{Lagos} C.~d.~P.,  {Lacey} C.~G.,   {Baugh} C.~M.,  2013, \mn@doi [\mnras]
  {10.1093/mnras/stt1696}, \href
  {http://adsabs.harvard.edu/abs/2013MNRAS.436.1787L} {436, 1787}

\bibitem[\protect\citeauthoryear{{Le Floc'h} et~al.,}{{Le Floc'h}
  et~al.}{2003}]{Lefloch03-2003A&A...400..499L}
{Le Floc'h} E.,  et~al., 2003, \mn@doi [\aap] {10.1051/0004-6361:20030001},
  \href {http://adsabs.harvard.edu/abs/2003A%26A...400..499L} {400, 499}

\bibitem[\protect\citeauthoryear{{Le Floc'h}, {Charmandaris}, {Forrest},
  {Mirabel}, {Armus}  \& {Devost}}{{Le Floc'h}
  et~al.}{2006}]{LeFloch06-2006ApJ...642..636L}
{Le Floc'h} E.,  {Charmandaris} V.,  {Forrest} W.~J.,  {Mirabel} I.~F.,
  {Armus} L.,   {Devost} D.,  2006, \mn@doi [\apj] {10.1086/501118}, \href
  {http://adsabs.harvard.edu/abs/2006ApJ...642..636L} {642, 636}

\bibitem[\protect\citeauthoryear{{Le Floc'h}, {Charmandaris}, {Gordon},
  {Forrest}, {Brandl}, {Schaerer}, {Dessauges-Zavadsky}  \& {Armus}}{{Le
  Floc'h} et~al.}{2012}]{LeFloch12-2012ApJ...746....7L}
{Le Floc'h} E.,  {Charmandaris} V.,  {Gordon} K.,  {Forrest} W.~J.,  {Brandl}
  B.,  {Schaerer} D.,  {Dessauges-Zavadsky} M.,   {Armus} L.,  2012, \mn@doi
  [\apj] {10.1088/0004-637X/746/1/7}, \href
  {http://adsabs.harvard.edu/abs/2012ApJ...746....7L} {746, 7}

\bibitem[\protect\citeauthoryear{{Leitherer} et~al.,}{{Leitherer}
  et~al.}{1999}]{Leitherer99-1999ApJS..123....3L}
{Leitherer} C.,  et~al., 1999, \mn@doi [\apjs] {10.1086/313233}, \href
  {http://adsabs.harvard.edu/abs/1999ApJS..123....3L} {123, 3}

\bibitem[\protect\citeauthoryear{{Leroy}, {Walter}, {Brinks}, {Bigiel}, {de
  Blok}, {Madore}  \& {Thornley}}{{Leroy}
  et~al.}{2008}]{Leroy08-2008AJ....136.2782L}
{Leroy} A.~K.,  {Walter} F.,  {Brinks} E.,  {Bigiel} F.,  {de Blok} W.~J.~G.,
  {Madore} B.,   {Thornley} M.~D.,  2008, \mn@doi [\aj]
  {10.1088/0004-6256/136/6/2782}, \href
  {http://adsabs.harvard.edu/abs/2008AJ....136.2782L} {136, 2782}

\bibitem[\protect\citeauthoryear{{Lyman} et~al.,}{{Lyman}
  et~al.}{2017}]{Lyman17-2017MNRAS.467.1795L}
{Lyman} J.~D.,  et~al., 2017, \mn@doi [\mnras] {10.1093/mnras/stx220}, \href
  {http://adsabs.harvard.edu/abs/2017MNRAS.467.1795L} {467, 1795}

\bibitem[\protect\citeauthoryear{{MacFadyen} \& {Woosley}}{{MacFadyen} \&
  {Woosley}}{1999}]{MacFadyen99-1999ApJ...524..262M}
{MacFadyen} A.~I.,  {Woosley} S.~E.,  1999, \mn@doi [\apj] {10.1086/307790},
  \href {http://adsabs.harvard.edu/abs/1999ApJ...524..262M} {524, 262}

\bibitem[\protect\citeauthoryear{{Malesani} et~al.,}{{Malesani}
  et~al.}{2004}]{Malesani04-2004ApJ...609L...5M}
{Malesani} D.,  et~al., 2004, \mn@doi [\apjl] {10.1086/422684}, \href
  {http://adsabs.harvard.edu/abs/2004ApJ...609L...5M} {609, L5}

\bibitem[\protect\citeauthoryear{{Marks}, {Kroupa}, {Dabringhausen}  \&
  {Pawlowski}}{{Marks} et~al.}{2012}]{Marks12-2012MNRAS.422.2246M}
{Marks} M.,  {Kroupa} P.,  {Dabringhausen} J.,   {Pawlowski} M.~S.,  2012,
  \mn@doi [\mnras] {10.1111/j.1365-2966.2012.20767.x}, \href
  {http://adsabs.harvard.edu/abs/2012MNRAS.422.2246M} {422, 2246}

\bibitem[\protect\citeauthoryear{{Mason}, {Hartkopf}, {Gies}, {Henry}  \&
  {Helsel}}{{Mason} et~al.}{2009}]{Mason09-2009AJ....137.3358M}
{Mason} B.~D.,  {Hartkopf} W.~I.,  {Gies} D.~R.,  {Henry} T.~J.,   {Helsel}
  J.~W.,  2009, \mn@doi [\aj] {10.1088/0004-6256/137/2/3358}, \href
  {http://adsabs.harvard.edu/abs/2009AJ....137.3358M} {137, 3358}

\bibitem[\protect\citeauthoryear{{McGaugh}}{{McGaugh}}{2012}]{McGaugh12-2012AJ....143...40M}
{McGaugh} S.~S.,  2012, \mn@doi [\aj] {10.1088/0004-6256/143/2/40}, \href
  {http://adsabs.harvard.edu/abs/2012AJ....143...40M} {143, 40}

\bibitem[\protect\citeauthoryear{{Micha{\l}owski} et~al.,}{{Micha{\l}owski}
  et~al.}{2014}]{Michalowski14-2014A&A...562A..70M}
{Micha{\l}owski} M.~J.,  et~al., 2014, \mn@doi [\aap]
  {10.1051/0004-6361/201322843}, \href
  {http://adsabs.harvard.edu/abs/2014A%26A...562A..70M} {562, A70}

\bibitem[\protect\citeauthoryear{{Paczy{\'n}ski}}{{Paczy{\'n}ski}}{1998}]{Paczynski98-1998ApJ...494L..45P}
{Paczy{\'n}ski} B.,  1998, \mn@doi [\apjl] {10.1086/311148}, \href
  {http://adsabs.harvard.edu/abs/1998ApJ...494L..45P} {494, L45}

\bibitem[\protect\citeauthoryear{{Pan} et~al.,}{{Pan}
  et~al.}{2018}]{Pan18-2018arXiv181010162P}
{Pan} H.-A.,  et~al., 2018, preprint, \href
  {http://adsabs.harvard.edu/abs/2018arXiv181010162P} {} (\mn@eprint {arXiv}
  {1810.10162})

\bibitem[\protect\citeauthoryear{{Peacock} et~al.,}{{Peacock}
  et~al.}{2017}]{Peacock17-2017ApJ...841...28P}
{Peacock} M.~B.,  et~al., 2017, \mn@doi [\apj] {10.3847/1538-4357/aa70eb},
  \href {http://adsabs.harvard.edu/abs/2017ApJ...841...28P} {841, 28}

\bibitem[\protect\citeauthoryear{{Pellerin}, {Meurer}, {Bekki}, {Elmegreen},
  {Wong}  \& {Knezek}}{{Pellerin}
  et~al.}{2010}]{Pellerin10-2010AJ....139.1369P}
{Pellerin} A.,  {Meurer} G.~R.,  {Bekki} K.,  {Elmegreen} D.~M.,  {Wong} O.~I.,
    {Knezek} P.~M.,  2010, \mn@doi [\aj] {10.1088/0004-6256/139/4/1369}, \href
  {http://cdsads.u-strasbg.fr/abs/2010AJ....139.1369P} {139, 1369}

\bibitem[\protect\citeauthoryear{{Perley} et~al.,}{{Perley}
  et~al.}{2013}]{Perley13-2013ApJ...778..128P}
{Perley} D.~A.,  et~al., 2013, \mn@doi [\apj] {10.1088/0004-637X/778/2/128},
  \href {http://adsabs.harvard.edu/abs/2013ApJ...778..128P} {778, 128}

\bibitem[\protect\citeauthoryear{{Perley} et~al.,}{{Perley}
  et~al.}{2015}]{Perley15-2015ApJ...801..102P}
{Perley} D.~A.,  et~al., 2015, \mn@doi [\apj] {10.1088/0004-637X/801/2/102},
  \href {http://adsabs.harvard.edu/abs/2015ApJ...801..102P} {801, 102}

\bibitem[\protect\citeauthoryear{{Perley} et~al.,}{{Perley}
  et~al.}{2016}]{Perley16-2016ApJ...817....8P}
{Perley} D.~A.,  et~al., 2016, \mn@doi [\apj] {10.3847/0004-637X/817/1/8},
  \href {http://adsabs.harvard.edu/abs/2016ApJ...817....8P} {817, 8}

\bibitem[\protect\citeauthoryear{{Pian} et~al.,}{{Pian}
  et~al.}{2006}]{Pian06-2006Natur.442.1011P}
{Pian} E.,  et~al., 2006, \mn@doi [\nat] {10.1038/nature05082}, \href
  {http://adsabs.harvard.edu/abs/2006Natur.442.1011P} {442, 1011}

\bibitem[\protect\citeauthoryear{{Piran}}{{Piran}}{2004}]{Piran04-2004RvMP...76.1143P}
{Piran} T.,  2004, \mn@doi [Reviews of Modern Physics]
  {10.1103/RevModPhys.76.1143}, \href
  {http://adsabs.harvard.edu/abs/2004RvMP...76.1143P} {76, 1143}

\bibitem[\protect\citeauthoryear{{Piran}, {Bromberg}, {Nakar}  \&
  {Sari}}{{Piran} et~al.}{2013}]{Piran13-2013RSPTA.37120273P}
{Piran} T.,  {Bromberg} O.,  {Nakar} E.,   {Sari} R.,  2013, \mn@doi
  [Philosophical Transactions of the Royal Society of London Series A]
  {10.1098/rsta.2012.0273}, \href
  {http://adsabs.harvard.edu/abs/2013RSPTA.37120273P} {371, 20120273}

\bibitem[\protect\citeauthoryear{{Podsiadlowski}, {Mazzali}, {Nomoto},
  {Lazzati}  \& {Cappellaro}}{{Podsiadlowski}
  et~al.}{2004}]{Podsiadlowski04-2004ApJ...607L..17P}
{Podsiadlowski} P.,  {Mazzali} P.~A.,  {Nomoto} K.,  {Lazzati} D.,
  {Cappellaro} E.,  2004, \mn@doi [\apjl] {10.1086/421347}, \href
  {http://adsabs.harvard.edu/abs/2004ApJ...607L..17P} {607, L17}

\bibitem[\protect\citeauthoryear{{Podsiadlowski}, {Ivanova}, {Justham}  \&
  {Rappaport}}{{Podsiadlowski}
  et~al.}{2010}]{Podsiadlowski10-2010MNRAS.406..840P}
{Podsiadlowski} P.,  {Ivanova} N.,  {Justham} S.,   {Rappaport} S.,  2010,
  \mn@doi [\mnras] {10.1111/j.1365-2966.2010.16751.x}, \href
  {http://adsabs.harvard.edu/abs/2010MNRAS.406..840P} {406, 840}

\bibitem[\protect\citeauthoryear{{Powell}, {Bournaud}, {Chapon}  \&
  {Teyssier}}{{Powell} et~al.}{2013}]{Powell13-2013MNRAS.434.1028P}
{Powell} L.~C.,  {Bournaud} F.,  {Chapon} D.,   {Teyssier} R.,  2013, \mn@doi
  [\mnras] {10.1093/mnras/stt1036}, \href
  {http://cdsads.u-strasbg.fr/abs/2013MNRAS.434.1028P} {434, 1028}

\bibitem[\protect\citeauthoryear{{Prochaska}, {Chen}, {Wolfe},
  {Dessauges-Zavadsky}  \& {Bloom}}{{Prochaska}
  et~al.}{2008}]{Prochaska08-2008ApJ...672...59P}
{Prochaska} J.~X.,  {Chen} H.-W.,  {Wolfe} A.~M.,  {Dessauges-Zavadsky} M.,
  {Bloom} J.~S.,  2008, \mn@doi [\apj] {10.1086/523689}, \href
  {http://adsabs.harvard.edu/abs/2008ApJ...672...59P} {672, 59}

\bibitem[\protect\citeauthoryear{{Renaud}}{{Renaud}}{2018}]{Renaud-2018NewAR..81....1R}
{Renaud} F.,  2018, \mn@doi [\nar] {10.1016/j.newar.2018.03.001}, \href
  {http://adsabs.harvard.edu/abs/2018NewAR..81....1R} {81, 1}

\bibitem[\protect\citeauthoryear{{Renaud}, {Bournaud}, {Kraljic}  \&
  {Duc}}{{Renaud} et~al.}{2014}]{Renaud14-2014MNRAS.442L..33R}
{Renaud} F.,  {Bournaud} F.,  {Kraljic} K.,   {Duc} P.-A.,  2014, \mn@doi
  [\mnras] {10.1093/mnrasl/slu050}, \href
  {http://adsabs.harvard.edu/abs/2014MNRAS.442L..33R} {442, L33}

\bibitem[\protect\citeauthoryear{{Renaud} et~al.,}{{Renaud}
  et~al.}{2018}]{Renaud18-2018MNRAS.473..585R}
{Renaud} F.,  et~al., 2018, \mn@doi [\mnras] {10.1093/mnras/stx2360}, \href
  {http://adsabs.harvard.edu/abs/2018MNRAS.473..585R} {473, 585}

\bibitem[\protect\citeauthoryear{{Saintonge} et~al.,}{{Saintonge}
  et~al.}{2011}]{Saintonge11-2011MNRAS.415...32S}
{Saintonge} A.,  et~al., 2011, \mn@doi [\mnras]
  {10.1111/j.1365-2966.2011.18677.x}, \href
  {http://adsabs.harvard.edu/abs/2011MNRAS.415...32S} {415, 32}

\bibitem[\protect\citeauthoryear{{Sana} et~al.,}{{Sana}
  et~al.}{2012}]{Sana12-2012Sci...337..444S}
{Sana} H.,  et~al., 2012, \mn@doi [Science] {10.1126/science.1223344}, \href
  {http://adsabs.harvard.edu/abs/2012Sci...337..444S} {337, 444}

\bibitem[\protect\citeauthoryear{{Sana} et~al.,}{{Sana}
  et~al.}{2014}]{Sana14-2014ApJS..215...15S}
{Sana} H.,  et~al., 2014, \mn@doi [\apjs] {10.1088/0067-0049/215/1/15}, \href
  {http://adsabs.harvard.edu/abs/2014ApJS..215...15S} {215, 15}

\bibitem[\protect\citeauthoryear{{Savaglio}}{{Savaglio}}{2015}]{Savaglio15-2015JHEAp...7...95S}
{Savaglio} S.,  2015, \mn@doi [Journal of High Energy Astrophysics]
  {10.1016/j.jheap.2015.06.004}, \href
  {http://adsabs.harvard.edu/abs/2015JHEAp...7...95S} {7, 95}

\bibitem[\protect\citeauthoryear{{Savaglio}, {Glazebrook}  \& {Le
  Borgne}}{{Savaglio} et~al.}{2009}]{Savaglio09-2009ApJ...691..182S}
{Savaglio} S.,  {Glazebrook} K.,   {Le Borgne} D.,  2009, \mn@doi [\apj]
  {10.1088/0004-637X/691/1/182}, \href
  {http://adsabs.harvard.edu/abs/2009ApJ...691..182S} {691, 182}

\bibitem[\protect\citeauthoryear{{Savaglio} et~al.,}{{Savaglio}
  et~al.}{2012}]{Savaglio12-2012MNRAS.420..627S}
{Savaglio} S.,  et~al., 2012, \mn@doi [\mnras]
  {10.1111/j.1365-2966.2011.20074.x}, \href
  {http://adsabs.harvard.edu/abs/2012MNRAS.420..627S} {420, 627}

\bibitem[\protect\citeauthoryear{{Schady} et~al.,}{{Schady}
  et~al.}{2015}]{Schady15-2015A&A...579A.126S}
{Schady} P.,  et~al., 2015, \mn@doi [\aap] {10.1051/0004-6361/201526060}, \href
  {http://adsabs.harvard.edu/abs/2015A%26A...579A.126S} {579, A126}

\bibitem[\protect\citeauthoryear{{Schaye} \& {Dalla Vecchia}}{{Schaye} \&
  {Dalla Vecchia}}{2008}]{Schaye08-2008MNRAS.383.1210S}
{Schaye} J.,  {Dalla Vecchia} C.,  2008, \mn@doi [\mnras]
  {10.1111/j.1365-2966.2007.12639.x}, \href
  {http://adsabs.harvard.edu/abs/2008MNRAS.383.1210S} {383, 1210}

\bibitem[\protect\citeauthoryear{{Schaye} et~al.,}{{Schaye}
  et~al.}{2010}]{Schaye10-2010MNRAS.402.1536S}
{Schaye} J.,  et~al., 2010, \mn@doi [\mnras]
  {10.1111/j.1365-2966.2009.16029.x}, \href
  {http://adsabs.harvard.edu/abs/2010MNRAS.402.1536S} {402, 1536}

\bibitem[\protect\citeauthoryear{{Schneider} et~al.,}{{Schneider}
  et~al.}{2018}]{Schneider17-2018Sci...359...69S}
{Schneider} F.~R.~N.,  et~al., 2018, \mn@doi [Science]
  {10.1126/science.aan0106}, \href
  {http://adsabs.harvard.edu/abs/2018Sci...359...69S} {359, 69}

\bibitem[\protect\citeauthoryear{{Sharma}, {Theuns}, {Frenk}, {Bower}, {Crain},
  {Schaller}  \& {Schaye}}{{Sharma}
  et~al.}{2017}]{Sharma17-2017MNRAS.468.2176S}
{Sharma} M.,  {Theuns} T.,  {Frenk} C.,  {Bower} R.~G.,  {Crain} R.~A.,
  {Schaller} M.,   {Schaye} J.,  2017, \mn@doi [\mnras] {10.1093/mnras/stx578},
  \href {http://adsabs.harvard.edu/abs/2017MNRAS.468.2176S} {468, 2176}

\bibitem[\protect\citeauthoryear{{Sokolov} et~al.,}{{Sokolov}
  et~al.}{2001}]{Sokolov01-2001A&A...372..438S}
{Sokolov} V.~V.,  et~al., 2001, \mn@doi [\aap] {10.1051/0004-6361:20010512},
  \href {http://adsabs.harvard.edu/abs/2001A%26A...372..438S} {372, 438}

\bibitem[\protect\citeauthoryear{{Sollerman}, {{\"O}stlin}, {Fynbo}, {Hjorth},
  {Fruchter}  \& {Pedersen}}{{Sollerman}
  et~al.}{2005}]{Sollerman05-2005NewA...11..103S}
{Sollerman} J.,  {{\"O}stlin} G.,  {Fynbo} J.~P.~U.,  {Hjorth} J.,  {Fruchter}
  A.,   {Pedersen} K.,  2005, \mn@doi [\na] {10.1016/j.newast.2005.06.004},
  \href {http://adsabs.harvard.edu/abs/2005NewA...11..103S} {11, 103}

\bibitem[\protect\citeauthoryear{{Springel}}{{Springel}}{2005}]{Springel05-2005MNRAS.364.1105S}
{Springel} V.,  2005, \mn@doi [\mnras] {10.1111/j.1365-2966.2005.09655.x},
  \href {http://adsabs.harvard.edu/abs/2005MNRAS.364.1105S} {364, 1105}

\bibitem[\protect\citeauthoryear{{Springel}, {Yoshida}  \& {White}}{{Springel}
  et~al.}{2001}]{Springel01-2001NewA....6...79S}
{Springel} V.,  {Yoshida} N.,   {White} S.~D.~M.,  2001, \mn@doi [\na]
  {10.1016/S1384-1076(01)00042-2}, \href
  {http://adsabs.harvard.edu/abs/2001NewA....6...79S} {6, 79}

\bibitem[\protect\citeauthoryear{{Stanek} et~al.,}{{Stanek}
  et~al.}{2003}]{Stanek03-2003ApJ...591L..17S}
{Stanek} K.~Z.,  et~al., 2003, \mn@doi [\apjl] {10.1086/376976}, \href
  {http://adsabs.harvard.edu/abs/2003ApJ...591L..17S} {591, L17}

\bibitem[\protect\citeauthoryear{{Starkenburg}, {Helmi}  \&
  {Sales}}{{Starkenburg} et~al.}{2016}]{Starkenburg16-2016A&A...587A..24S}
{Starkenburg} T.~K.,  {Helmi} A.,   {Sales} L.~V.,  2016, \mn@doi [\aap]
  {10.1051/0004-6361/201527247}, \href
  {http://adsabs.harvard.edu/abs/2016A%26A...587A..24S} {587, A24}

\bibitem[\protect\citeauthoryear{{Struck}, {Appleton}, {Borne}  \&
  {Lucas}}{{Struck} et~al.}{1996}]{Struck96-1996AJ....112.1868S}
{Struck} C.,  {Appleton} P.~N.,  {Borne} K.~D.,   {Lucas} R.~A.,  1996, \mn@doi
  [\aj] {10.1086/118148}, \href
  {http://cdsads.u-strasbg.fr/abs/1996AJ....112.1868S} {112, 1868}

\bibitem[\protect\citeauthoryear{{Svensson}, {Levan}, {Tanvir}, {Fruchter}  \&
  {Strolger}}{{Svensson} et~al.}{2010}]{Svensson10-2010MNRAS.405...57S}
{Svensson} K.~M.,  {Levan} A.~J.,  {Tanvir} N.~R.,  {Fruchter} A.~S.,
  {Strolger} L.-G.,  2010, \mn@doi [\mnras] {10.1111/j.1365-2966.2010.16442.x},
  \href {http://adsabs.harvard.edu/abs/2010MNRAS.405...57S} {405, 57}

\bibitem[\protect\citeauthoryear{{Svensson} et~al.,}{{Svensson}
  et~al.}{2012}]{Svensson12-2012MNRAS.421...25S}
{Svensson} K.~M.,  et~al., 2012, \mn@doi [\mnras]
  {10.1111/j.1365-2966.2011.19811.x}, \href
  {http://adsabs.harvard.edu/abs/2012MNRAS.421...25S} {421, 25}

\bibitem[\protect\citeauthoryear{{Tanvir} et~al.,}{{Tanvir}
  et~al.}{2009}]{Tanvir09}
{Tanvir} N.~R.,  et~al., 2009, \mn@doi [\nat] {10.1038/nature08459}, \href
  {http://adsabs.harvard.edu/abs/2009Natur.461.1254T} {461, 1254}

\bibitem[\protect\citeauthoryear{{Teyssier}, {Chapon}  \&
  {Bournaud}}{{Teyssier} et~al.}{2010}]{Teyssier10-2010ApJ...720L.149T}
{Teyssier} R.,  {Chapon} D.,   {Bournaud} F.,  2010, \mn@doi [\apjl]
  {10.1088/2041-8205/720/2/L149}, \href
  {http://adsabs.harvard.edu/abs/2010ApJ...720L.149T} {720, L149}

\bibitem[\protect\citeauthoryear{{Tout}, {Wickramasinghe}, {Lau}, {Pringle}  \&
  {Ferrario}}{{Tout} et~al.}{2011}]{Tout11-2011MNRAS.410.2458T}
{Tout} C.~A.,  {Wickramasinghe} D.~T.,  {Lau} H.~H.-B.,  {Pringle} J.~E.,
  {Ferrario} L.,  2011, \mn@doi [\mnras] {10.1111/j.1365-2966.2010.17622.x},
  \href {http://adsabs.harvard.edu/abs/2011MNRAS.410.2458T} {410, 2458}

\bibitem[\protect\citeauthoryear{{Usov}}{{Usov}}{1992}]{Usov92-1992Natur.357..472U}
{Usov} V.~V.,  1992, \mn@doi [\nat] {10.1038/357472a0}, \href
  {http://adsabs.harvard.edu/abs/1992Natur.357..472U} {357, 472}

\bibitem[\protect\citeauthoryear{{Wainwright}, {Berger}  \&
  {Penprase}}{{Wainwright} et~al.}{2007}]{Wainwright07-2007ApJ...657..367W}
{Wainwright} C.,  {Berger} E.,   {Penprase} B.~E.,  2007, \mn@doi [\apj]
  {10.1086/510794}, \href {http://adsabs.harvard.edu/abs/2007ApJ...657..367W}
  {657, 367}

\bibitem[\protect\citeauthoryear{{Weidner}, {Bonnell}  \&
  {Zinnecker}}{{Weidner} et~al.}{2010}]{Weidner10-2010ApJ...724.1503W}
{Weidner} C.,  {Bonnell} I.~A.,   {Zinnecker} H.,  2010, \mn@doi [\apj]
  {10.1088/0004-637X/724/2/1503}, \href
  {http://adsabs.harvard.edu/abs/2010ApJ...724.1503W} {724, 1503}

\bibitem[\protect\citeauthoryear{{Wiseman}, {Perley}, {Schady}, {Prochaska},
  {de Ugarte Postigo}, {Kr{\"u}hler}, {Yates}  \& {Greiner}}{{Wiseman}
  et~al.}{2017}]{Wiseman17-2017A&A...607A.107W}
{Wiseman} P.,  {Perley} D.~A.,  {Schady} P.,  {Prochaska} J.~X.,  {de Ugarte
  Postigo} A.,  {Kr{\"u}hler} T.,  {Yates} R.~M.,   {Greiner} J.,  2017,
  \mn@doi [\aap] {10.1051/0004-6361/201731065}, \href
  {http://adsabs.harvard.edu/abs/2017A%26A...607A.107W} {607, A107}

\bibitem[\protect\citeauthoryear{{Wong} et~al.,}{{Wong}
  et~al.}{2006}]{Wong06-2006MNRAS.370.1607W}
{Wong} O.~I.,  et~al., 2006, \mn@doi [\mnras]
  {10.1111/j.1365-2966.2006.10589.x}, \href
  {http://adsabs.harvard.edu/abs/2006MNRAS.370.1607W} {370, 1607}

\bibitem[\protect\citeauthoryear{{Woosley}}{{Woosley}}{1993}]{Woosley93-1993ApJ...405..273W}
{Woosley} S.~E.,  1993, \mn@doi [\apj] {10.1086/172359}, \href
  {http://adsabs.harvard.edu/abs/1993ApJ...405..273W} {405, 273}

\bibitem[\protect\citeauthoryear{{Woosley} \& {Heger}}{{Woosley} \&
  {Heger}}{2006}]{Woosley06-2006ApJ...637..914W}
{Woosley} S.~E.,  {Heger} A.,  2006, \mn@doi [\apj] {10.1086/498500}, \href
  {http://adsabs.harvard.edu/abs/2006ApJ...637..914W} {637, 914}

\bibitem[\protect\citeauthoryear{{Yoon} \& {Langer}}{{Yoon} \&
  {Langer}}{2005}]{Yoon05-2005A&A...443..643Y}
{Yoon} S.-C.,  {Langer} N.,  2005, \mn@doi [\aap] {10.1051/0004-6361:20054030},
  \href {http://adsabs.harvard.edu/abs/2005A%26A...443..643Y} {443, 643}

\bibitem[\protect\citeauthoryear{{Zhang}, {Woosley}  \& {MacFadyen}}{{Zhang}
  et~al.}{2003}]{Zhang03-2003ApJ...586..356Z}
{Zhang} W.,  {Woosley} S.~E.,   {MacFadyen} A.~I.,  2003, \mn@doi [\apj]
  {10.1086/367609}, \href {http://adsabs.harvard.edu/abs/2003ApJ...586..356Z}
  {586, 356}

\bibitem[\protect\citeauthoryear{{Zhang}, {Woosley}  \& {Heger}}{{Zhang}
  et~al.}{2004}]{Zhang04-2004ApJ...608..365Z}
{Zhang} W.,  {Woosley} S.~E.,   {Heger} A.,  2004, \mn@doi [\apj]
  {10.1086/386300}, \href {http://adsabs.harvard.edu/abs/2004ApJ...608..365Z}
  {608, 365}

\bibitem[\protect\citeauthoryear{{Zhang}, {Romano}, {Ivison}, {Papadopoulos}
  \& {Matteucci}}{{Zhang} et~al.}{2018}]{Zhang18-2018Natur.558..260Z}
{Zhang} Z.-Y.,  {Romano} D.,  {Ivison} R.~J.,  {Papadopoulos} P.~P.,
  {Matteucci} F.,  2018, \mn@doi [\nat] {10.1038/s41586-018-0196-x}, \href
  {http://adsabs.harvard.edu/abs/2018Natur.558..260Z} {558, 260}

\bibitem[\protect\citeauthoryear{{Zhu}, {Wu}, {Li}  \& {Cao}}{{Zhu}
  et~al.}{2010}]{Zhu10-2010RAA....10..329Z}
{Zhu} Y.-N.,  {Wu} H.,  {Li} H.-N.,   {Cao} C.,  2010, \mn@doi [Research in
  Astronomy and Astrophysics] {10.1088/1674-4527/10/4/004}, \href
  {http://adsabs.harvard.edu/abs/2010RAA....10..329Z} {10, 329}

\bibitem[\protect\citeauthoryear{{de Grijs}, {Lee}, {Clemencia Mora Herrera},
  {Fritze-v.~Alvensleben}  \& {Anders}}{{de Grijs}
  et~al.}{2003}]{deGrijs03-2003NewA....8..155D}
{de Grijs} R.,  {Lee} J.~T.,  {Clemencia Mora Herrera} M.,
  {Fritze-v.~Alvensleben} U.,   {Anders} P.,  2003, \mn@doi [\na]
  {10.1016/S1384-1076(02)00224-5}, \href
  {http://adsabs.harvard.edu/abs/2003NewA....8..155D} {8, 155}

\bibitem[\protect\citeauthoryear{{van den Heuvel} \& {Portegies Zwart}}{{van
  den Heuvel} \& {Portegies Zwart}}{2013}]{Heuvel13-2013ApJ...779..114V}
{van den Heuvel} E.~P.~J.,  {Portegies Zwart} S.~F.,  2013, \mn@doi [\apj]
  {10.1088/0004-637X/779/2/114}, \href
  {http://adsabs.harvard.edu/abs/2013ApJ...779..114V} {779, 114}

\makeatother
\end{thebibliography}

\bsp

\label{lastpage}

\end{document}